\documentclass[sigconf, screen]{acmart}

\AtBeginDocument{%
  \providecommand\BibTeX{{%
    \normalfont B\kern-0.5em{\scshape i\kern-0.25em b}\kern-0.8em\TeX}}}

\setcopyright{acmcopyright}
\acmPrice{15.00}
\acmDOI{10.1145/3368089.3409717}
\acmYear{2020}
\copyrightyear{2020}
\acmSubmissionID{fse20main-p383-p}
\acmISBN{978-1-4503-7043-1/20/11}
\acmConference[ESEC/FSE '20]{Proceedings of the 28th ACM Joint European Software Engineering Conference and Symposium on the Foundations of Software Engineering}{November 8--13, 2020}{Virtual Event, USA}
\acmBooktitle{Proceedings of the 28th ACM Joint European Software Engineering Conference and Symposium on the Foundations of Software Engineering (ESEC/FSE '20), November 8--13, 2020, Virtual Event, USA}

\usepackage{tabularx}
\usepackage{booktabs}
\usepackage{float}
\usepackage{graphbox}
\usepackage{cite}
\usepackage{amsmath,amsfonts}
\usepackage{algorithmic}
\usepackage{graphicx}
\usepackage{textcomp}
\usepackage{xcolor}
\usepackage{dcolumn,booktabs}
\usepackage{listings}
\usepackage{wrapfig}
\usepackage{dcolumn}
\usepackage{verbatim}
\usepackage{fontawesome}
\usepackage{marvosym}
\usepackage{natbib}
\usepackage{framed}
\usepackage{xspace}

\usepackage{microtype} 

\newcolumntype{d}[1]{D{.}{.}{#1}}

\newcommand{\ING}{ING\xspace}

\def\BibTeX{{\rm B\kern-.05em{\sc i\kern-.025em b}\kern-.08em
    T\kern-.1667em\lower.7ex\hbox{E}\kern-.125emX}}

\begin{document}

\title{Questions for Data Scientists in Software Engineering: \\A Replication}

\author{Hennie Huijgens}
\affiliation{%
  \institution{Delft University of Technology}
  \city{Delft}
  \country{The Netherlands}}
\email{h.k.m.huijgens@tudelft.nl}

\author{Ayushi Rastogi}
\author{Ernst Mulders}
\authornote{Work completed during an internship at \ING.}
\affiliation{%
  \institution{Delft University of Technology}
  \city{Delft}
  \country{The Netherlands}}
\email{a.rastogi@tudelft.nl}
\email{ernst@mulde.rs}

\author{Georgios Gousios}
\author{Arie van Deursen}
\affiliation{%
  \institution{Delft University of Technology}
  \city{Delft}
  \country{The Netherlands}}
\email{g.gousios@tudelft.nl}
\email{arie.vandeursen@tudelft.nl}

\renewcommand{\shorttitle}{Questions for Data Scientists in Software Engineering: A Replication}

\begin{abstract}
In 2014, a Microsoft study investigated the sort of questions that data science applied to software engineering should answer.
This resulted in 145 questions that developers considered  relevant for data scientists to answer, thus providing a research agenda to the community.
Fast forward to five years, no further studies investigated whether the questions from the software engineers at Microsoft hold for other software companies, including software-intensive companies with different primary focus (to which we refer as \emph{software-defined enterprises}).
Furthermore, it is not evident that the problems identified five years ago are still applicable, given the technological advances in software engineering.

This paper presents a study at \ING, a
software-defined enterprise in banking in which over 15,000 IT staff provides in-house software solutions.
This paper presents a comprehensive guide of questions for data scientists selected from the previous study at Microsoft along with our current work at \ING. 
We replicated the original Microsoft study at \ING, looking for questions that impact both software companies and software-defined enterprises and continue to impact software engineering.
We also add new questions that emerged from differences in the context of the two companies and the five years gap in between.
Our results show that software engineering questions for data scientists in the software-defined enterprise are largely similar to the software company, albeit with exceptions.
We hope that the software engineering research community builds on the new list of questions to create a useful body of knowledge. 
\end{abstract}



\maketitle

\section{Introduction}
Software engineering researchers try solving problems that are relevant to software developers, teams, and organizations.
Historically, researchers identified these problems from their experience, connections in industry and/or prior research. 
In 2014, however, a study at Microsoft~\citep{Begel:2014:ATQ:2568225.2568233} systematically analyzed software engineering questions that data scientists can answer and made it accessible to a wider audience. 

Switching context, in the past few years \ING transformed itself from a finance-oriented company to a software-defined, data-driven enterprise.
From a software engineering perspective, this includes the implementation of fully automated release engineering pipelines for software development activities in more than 600 teams performing 2,500+ deployments per month for 750+ applications.
These activities leave a trove of data, suggesting that data scientists using, e.g., modern machine learning techniques could offer valuable and actionable insights to \ING.

To that end, \ING needs questions that are relevant for their engineers which their data scientists can answer. 
As we started looking for existing resources, we came across the 145 software engineering questions for data scientists presented in the Microsoft study~\citep{Begel:2014:ATQ:2568225.2568233}.
However, before adopting the list,
we wanted to know:

\begin{flushleft}
\textit{RQ:
To what extent do software engineering questions relevant for Microsoft apply to \ING, five years later?}
\end{flushleft}

\begin{figure}[t]
\centering
\includegraphics[width=0.48\textwidth]{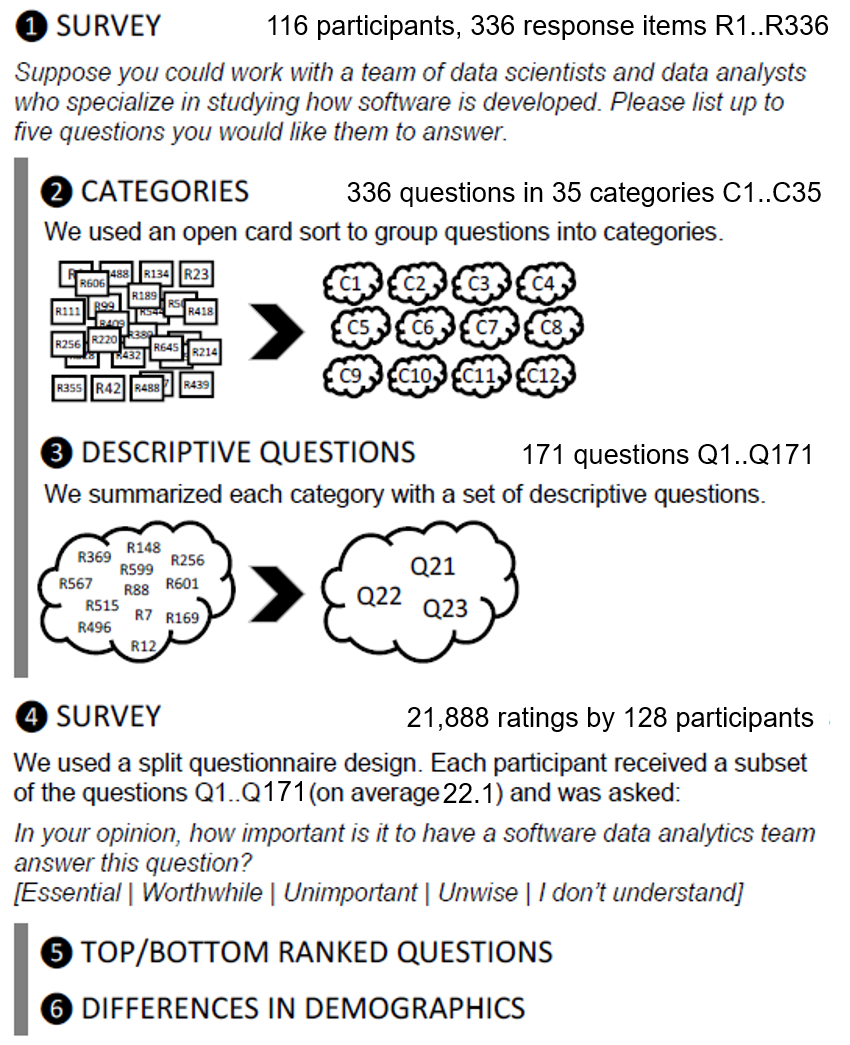}
\begin{flushleft}
{\footnotesize This figure is a copy from the original Microsoft 2014 study, with numbers from our study. The figure was re-used with permission of the Microsoft 2014 study authors.}
\end{flushleft}
\caption{Overview of the research methodology}
\label{fig:Research_approach}
\end{figure}

Microsoft is a large software company, while \ING that is a FinTech company using software to improve its banking solutions (software-defined enterprise).
Moreover, the two companies are at different scale.
In 2014, Microsoft had more than 30,000 engineers while even today \ING is almost half its size with approximately 15,000 IT employees (on a total of 45,000).
More details on the differences in the context of the two companies are available in Table~\ref{tab:Context}. 
We try to understand whether the questions relevant for a software company extend to a software-defined enterprise.
We compare the results of the original Microsoft study~\citep{Begel:2014:ATQ:2568225.2568233} with our results at \ING to understand the relevance of the questions beyond Microsoft but also as a guide for other software-defined enterprises that are undergoing their digital transformation.
We further explore whether the technological advances in the last five years changed the way we develop software.
To answer this question, we carried out a replication of the original Microsoft study at \ING.
Similar to the original study, we conducted two surveys: one, to find data science problems in software engineering, and second, to rank the questions in the order of their relevance (see Figure \ref{fig:Research_approach}).
For the first survey, we randomly sampled 1,002 \ING engineers and received 116 responses with 336 questions.
We grouped the 336 questions on similarities resulting in 171 descriptive questions.
We shared subsets of these 171 descriptive questions with another random sample of 1,296 \ING engineers for ranking.
In the end, we received 21,888 rankings from 128 \ING engineers.
These ranked 171 questions are the questions that engineers at \ING would like data scientists to solve. 
Further, we compare our list of 171 questions to the original list of 145 questions to answer our research question.
Our study shows that the core software development problems, relating to \emph{code} (e.g. understanding code, testing, and quality), \emph{developer productivity} (both individuals and team) and \emph{customer} are same for the software company and the software-defined enterprise.
Nonetheless, subtle differences in the type of questions point to changes in market as well as differences in the context of the two companies.

\begin{table}[b]
    \footnotesize
    \caption{Context of Microsoft in 2014 and \ING in 2019.}
    \label{tab:Context}
    \begin{tabularx}{\linewidth}{
    >{\hsize=0.44\hsize}X
    >{\hsize=0.5\hsize}X
    >{\hsize=0.5\hsize}X
    }
    \toprule
    & Microsoft 2014 & \ING 2019 \\
    \midrule
    Branch & Software Company & Banking (FinTech) \\
    Organization Size & Approx. 100,000 (in 2014), about 30,000 engineers & 45,000 employees of which 15,000 IT \\
    Team Structure & Typically size 5 $\pm$ 2  & 600 teams of size 9 $\pm$ 2 \\
    Development Model & Agile/Scrum (60\%+) & Agile (Scrum / Kanban) \\
    Pipeline automation & Every team is different. Continuous Integration in many teams & Continuous Delivery as a Service \\
    Development Practice & DevOps & (Biz)DevOps \\
    \addlinespace[1pt]
    \bottomrule
    \\[-4pt]
    \end{tabularx}
\end{table}




\section{Impact of the Microsoft 2014 Study}
In order to gain a good insight into the further course of the Microsoft 2014 study after it was published, including any implications for research, we conducted a citation analysis. In addition, we looked at studies that have not quoted the Microsoft study, but that are relevant to our study. Hence this section also serves as our discussion of related work.
We investigated the 136 studies that, according to Google Scholar, quote the Microsoft study. First of all, we looked at the number of times that the 136 studies themselves were cited by other studies; we limited the further analysis to 70 studies with a citation per year greater than 1.00. We then characterized studies into empirical approach, reference characterization, SE topic, and machine learning (ML) topic (see Table \ref{tab:Citing_empirical_approach}).
Note that one paper can belong to multiple topics. We made the following observations:

\paragraph{Microsoft itself is building on its study}
11\% of the citations come from Microsoft studies itself, mostly highly cited studies on SE culture, such as \citep{Kim:2016:ERD:2884781.2884783, 7886896, lo2015practitioners}. we notice that all citing Microsoft studies use a survey among a large number of SE practitioners (ranging from 16 to 793 respondents with a median of 311), whereas other studies based on a survey generally reach substantially lower numbers of participants.

\begin{table}[t]
    \footnotesize
    \caption{Characterizations of Citing Studies.}
    \label{tab:Citing_empirical_approach}
    \begin{tabular}{p{4.4cm}p{1.8cm}p{1.2cm}}
    \toprule
    Empirical Approach (n = 70) & Number of studies & Percentage \\
    \midrule
    Analysis of SE process data (e.g. IDE) & 30 & 43\% \\
    Survey SE practitioners & 17 & 24\% \\
    Interview SE practitioners & 7 & 10\% \\
    Literature review & 5 & 7\% \\
    Experiment, case, or field study & 5 & 7\% \\
    \addlinespace[1pt]
    \bottomrule
    \\[-4pt]
    \end{tabular}
\end{table}

\begin{table}[t]
    \footnotesize
    \label{tab:Citing_reference_characterization}
    \begin{tabular}{p{4.4cm}p{1.8cm}p{1.2cm}}
    \toprule
    Reference characterization (n = 70) & Number of studies & Percentage \\
    \midrule
    Plain reference in related work & 38 & 54\% \\
    Reference as example for study setup & 27 & 39\% \\
    Study partly answers MS question & 9 & 13\% \\
    Study explicitly answers MS question & 3 & 4\% \\
    \addlinespace[1pt]
    \bottomrule
    \\[-4pt]
    \end{tabular}
\end{table}

\begin{table}[t]
    \footnotesize
    \label{tab:Citing_SE_topic}
    \begin{tabular}{p{4.4cm}p{1.8cm}p{1.2cm}}
    \toprule
    Software Engineering Topic (n = 70) & Number of studies & Percentage \\
    \midrule
    Software analytics, data science & 20 & 29\% \\
    Testing, debugging, quality, code review & 15 & 21\% \\
    Software engineering process & 12 & 17\% \\
    Software engineering culture & 9 & 13\% \\
    Mobile apps & 3 & 4\% \\
    \addlinespace[1pt]
    \bottomrule
    \\[-4pt]
    \end{tabular}
\end{table}

\begin{table}[t]
    \footnotesize
    \label{tab:Citing_ML4SE_topic}
    \begin{tabular}{p{4.4cm}p{1.8cm}p{1.2cm}}
    \toprule
    Machine Learning Topic (n = 24) & Number of studies & Percentage \\
    \midrule
    Examples of Machine Learning applications & 8 & 11\%\\
    Natural Language Processing & 5 & 7\%\\
    Ensemble Algorithms & 3 & 4\%\\
    Instance-based Algorithms & 2 & 3\%\\
    Deep Learning Algorithms & 2 & 3\% \\
    Other & 4 & 5\%\\
    \addlinespace[1pt]
    \bottomrule
    \\[-4pt]
    \end{tabular}
\end{table}

\paragraph{Half of the citing studies analyze SE process data, and 24\% uses a survey}
Looking at the empirical approach (see the first sub-table in Table~\ref{tab:Citing_empirical_approach}), indicates that 43\% of the studies contain a quantitative component, in which analysis of SE process data in particular is part of the study. Good examples are \citep{beller2015and, gu2015parts}. Furthermore, 24\% of the citing studies uses a survey among SE practitioners, for example \citep{kochhar2016practitioners, 7886896, storey2016social, garousi2015survey, wan2018perceptions}. Ten percent is based on interviews with SE practitioners, such as \citep{kim2016emerging, ford2016paradise, Kim:2016:ERD:2884781.2884783, li2015makes}. Seven percent contains a literature review, for example \citep{kochhar2016practitioners, cartaxo2018role, tripathi2015university}. Another 7\% conducts an experiment \citep{roehm2015two, hilton2016tddviz}, case study \citep{leitner2016modelling, nayebi2019longitudinal}, or field study \citep{beller2015and, beller2015much}.

\paragraph{Only three out of 70 studies explicitly answer a question from the initial Microsoft study}
The second sub-table in Table \ref{tab:Citing_empirical_approach} shows that only 3 studies (4\%) explicitly refer their research question to an initial Microsoft one: \citep{gu2015parts, deewattananon2017analyzing, hilton2016tddviz}. Nine studies (13\%) partly try to answer a MS question: \citep{guzman2015ensemble, sawant2016reaction, sawant2018reaction, roehm2015two, lopez2014machine, suonsyrja2016post, beller2015much, beller2015and, beller2017developer}.
29 studies (39\%) refer to the original Microsoft study because they used it as an example for their own study \citep{nayebi2019longitudinal, denny2019research}, either with regard to the study design \citep{sharma2017developers, kabeer2017predicting, ford2016paradise, krishna2018bellwethers, garousi2015survey, kononenko2016code, Gupta:2015:ISP:2820518.2820560}, the rating approach (Kano) \citep{nayebi2017crowdsourced, lo2015practitioners}, or the card sorting technique \citep{mathis2017detecting, nayebi2018anatomy, ford2015exploring, sarkar2017characterizing}.
Furthermore, a large part (38 studies, 54\%) of the citing studies simply refers to the original Microsoft study in a simple related work way.

\paragraph{A majority of citing studies is about Software Analytics, Testing related studies, and SE Process}
The third sub-table shows that most cited studies are about software analytics, often combined with a focus on the role of the software engineer and its perceptions, e.g. \citep{kim2016emerging, lo2015practitioners}. In other cases the emphasis on software analytics is combined with a more technical focus on machine learning, e.g. \citep{fu2017easy, krishna2018bellwethers}. Other studies within the topic software analytics are about a variety of methods, tools, and techniques \citep{Bird20151, menzies2014sharing, andrei2016probabilistic, tripathi2015university, gousios2016streaming, suonsyrja2015designing, kafer2017summarizing, chen2018applications, krishna2017learning, arndt2018big, dkabrowski2019finding, tavakoli2018extracting}. 
Many of the studies that cite the Microsoft study---and which are often quoted themselves---relate to testing or test automation. 
Fifteen studies (21\%) are about testing \citep{beller2015and, hilton2016tddviz, kochhar2016practitioners, beller2015much, garousi2017worlds, schermann2016bifrost, garousi2016selecting, carver2016practitioners, beller2017developer}, debugging \citep{zou2018practitioners} and code review \citep{kononenko2016code, german2018my}.

12 studies (17\%) handle SE process related topics, such as productivity of software engineers \citep{lopez2014machine}, visualization \citep{hahn2015thread, batch2017interactive}, and continuous delivery \citep{vassallo2016continuous, widder2019conceptual}. 
In addition, studies also relate to continuous delivery pipelines and pipeline automation \citep{vassallo2016continuous, zampetti2017open}.
Another frequent topic in citing studies is data and models, including aspects of cloud development \citep{menzies2014sharing, leitner2016modelling, hassan2010software}.
Driven by a tendency toward automation of pipelines, software generates a large amount of data. Many different data sources---such as version control systems, peer code review systems, issue tracking systems, mail archives---are available for mining purposes \citep{Zimmermann:2017:SPD:3084100.3087674, Gupta:2015:ISP:2820518.2820560}. 

\paragraph{34\% of the cited studies includes some form of Machine Learning}
One third of the citing papers do include some form of machine learning (ML), ranging from applying a ML technique for analysis purposes to coming up with examples of the application of ML in practice. As the fourth sub-table in Table \ref{tab:Citing_empirical_approach} shows, 8 studies include examples of applications of ML in practice, e.g. \citep{Kim:2016:ERD:2884781.2884783, Bird20151, menzies2014sharing}. Text related techniques such as NLP occur 5 times, e.g. \citep{nayebi2017crowdsourced, garousi2017worlds}, ensemble techniques 3 times \citep{guzman2015ensemble, kabeer2017predicting, nayebi2018anatomy}, and instance-based and deep learning both 2 times \citep{gousios2016streaming, chen2018applications, fu2017easy, krishna2018bellwethers}. Four other techniques---neural networks, clustering, decision trees, and regression---occur one time. Perhaps this finding supports a trend that is visible in SE research, where more and more machine learning techniques are being used in SE analyzes and vice versa, also called \textit{AI-for-Software-Engineering} \citep{amershi2019software, khomh2018software, lwakatare2019taxonomy}.

\paragraph{13\% are about the cultural aspects of software engineering}
Software analytics is an area of extensive growth \citep{menzies2018software}. The original Microsoft 2014 study influenced ongoing research, looking at the 136 papers citing it gives the impression that it certainly did inspire other researchers and practitioners in setting up studies on software developers needs.
Nine studies (13\%) of the citing studies are about cultural aspects of software engineering, such as topic selection in experiments \citep{misirli2014topic}, characteristics of software engineers \citep{li2015makes, ford2016paradise, sharma2017developers}, causes for frustration \citep{ford2015exploring}, or challenges for software engineers \citep{Gupta:2015:ISP:2820518.2820560, storey2016social, sarkar2017characterizing}.

\section{Study Design}
Our study design comprises of two parts. 
In part one, we replicate the original Microsoft study at \ING.
We follow the step-by-step procedure prescribed in the original study, with slight modifications appropriate for our context
Figure~\ref{fig:Research_approach} depicts the research methodology we followed; the figure is an exact copy of the approach used in the original Microsoft 2014 study with numbers from our study. 
In the next step, we compare the questions identified in the Microsoft study to ours for similarities and differences including addition of new questions and removal of previous questions to answer our research questions.

\subsection{The Initial Survey}
We sent the initial survey to 1,002 \ING software engineers randomly chosen from a group of 2,342 employees working within the IT department of \ING in May 2018.
Unlike the Microsoft study, we did not offer any reward to increase the participation.
This is a deviation from the original study but aligns with the policy of \ING.
Out of the 1,002 engineers 387 engineers started the survey, 271 of them even filled the demographics but stopped when asked to write questions. 
In the end, we received 336 questions from 116 responses for a response rate of 11.6\%. 
Table \ref{tab:Area} shows the distribution of responses across discipline and role.  

\begin{table}[b]
    \footnotesize
    \caption{Distribution of responses based on discipline and role in the initial survey as well as rating survey.}
    \label{tab:Area}
    \begin{tabularx}{\linewidth}{
    >{\hsize=0.6\hsize}X
    >{\hsize=0.3\hsize}X
    >{\hsize=0.3\hsize}X
    }
    \toprule
    Discipline & Initial Survey & Rating Survey \\
    \midrule
    Development \& Testing & 62.0\% & 68.8\% \\
    Project Management & 2.0\% & 3.9\% \\
    Other Engineering (e.g. architect) & 28.0\% & 19.5\% \\
    Non-Engineering & 8.0\% & 7.8\% \\
    \addlinespace[1pt]
    \bottomrule
    \\[-4pt]
    \toprule
    Current Role & Initial Survey & Rating Survey \\
    \midrule
    Developer & 51.1\% & 20.0\% \\
    Lead & 14.3\% & 18.7\% \\
    Architect & 9.0\% & 11.8\% \\
    Manager \& Executive & 8.3\% & 20.0\% \\
    Other & 17.3\% & 29.6\% \\
    \addlinespace[1pt]
    \bottomrule
    \\[-4pt]
    \end{tabularx}
\end{table}

\subsection{Coding and Categorization}
Next we did an open card sort to group 336 questions into categories.
Our card sort was open, meaning that we coded independently from the Microsoft study. 
To create independent codes, the first author who did a majority of the coding did not study the Microsoft paper before or during the replication.
The other authors knew the paper from before and merely skimmed the methodology section for replication.

We let the groups emerge and evolve during the sorting process.
This process comprised of three phases.
In \textit{preparation phase}, we created a card for each question. 
Questions 1~to~40 were tagged by the second author. 
Questions 41~to~80 were tagged by the fourth author. 
Questions 81~to~90 were tagged by both the second and the fourth author. 
The tags of questions 1~to~90 were discussed by both the second and fourth author and based on their discussion final tags were prepared. 
The remaining questions 91~to~336 were then tagged by the first author, based on the tags from the previous step. 
We discarded cards that made general comments on software development and did not inquire any specific topic.

In the \textit{execution phase}, cards were sorted into meaningful groups and were assigned a descriptive title.
Similar to the Microsoft study, the questions were not easy to work with; many questions were same or similar to one another, most were quite verbose while others were overly specific. We distilled them into a set of so-called descriptive questions that more concisely describe each category (and sub-category).
In this step, out of the 336 questions, 49 questions were discarded and the remaining 287 questions were divided into 35 sub-categories. 
An example of reaching descriptive question is presented below\footnote{A closed balloon indicates a respondent question; an open balloon indicates a descriptive question.}:

\vspace{0.5mm} \noindent \faCommentO \textit{ `What factors affect the composition of DevOps teams?' }

from the following respondents' questions:

\vspace{0.5mm} \noindent \faComment \textit{``Would it be better to create specialized development teams instead of DevOps teams?"}

\vspace{0.5mm} \noindent \faComment \textit{"What is your idea of an ideal team that should develop software? How many and what kind of people should be part of it?"}

Finally, in the \textit{analysis phase}, we created abstract hierarchies to deduce general categories and themes.
In total, we created 171 descriptive questions, a full list of which is available in the appendix.

\subsection{The Rating Survey}
We created a second survey to rate the 171 descriptive questions. 
We split the questionnaire into eight component blocks (similar to the Microsoft study) and sent component blocks to potential respondents.
The idea behind using the \textit{split questionnaire survey design} is to avoid low response rate.
Each participant received a block of questions along with a text \textit{"In your opinion, how important is it to have a software data analytics team answer this question?"} with possible answers as \textit{"Essential"}, \textit{"Worthwhile"}, \textit{"Unimportant"}, \textit{"Unwise"}, and \textit{"I don't understand"} \citep{Kano1984AttractiveQA}. 

The rating survey was sent to the remaining 1,296 software engineers at \ING. 
Here too, 360 engineers started the survey (28\%), but many of them did not complete it (36\% drop-out rate). 
Finally, we received 128 responses, for a somewhat low response rate of 10\%. 
On an average each question received 21,888/177=123 ratings making the resulting ranks stable.
Table \ref{tab:Area} shows the distribution of responses for the rating survey based on discipline and current role.

\subsubsection{Top-Rated/Bottom-Rated Questions}
Finally, to rank each question, we dichotomized the ordinal Kano scale avoiding any scale violations \citep{KitchenhamPfleeger2008}. 
We computed the following percentages for each descriptive question:

\begin{itemize}
    \item Percentage of '\textbf{Essential}' responses among all the responses:
        \[\frac{Essential}{Essential + Worthwhile + Unimportant + Unwise}\]
    \item Percentage of 'Essential' and 'Worthwhile' responses among all the responses (to which we refer as \textbf{Worthwhile+}):
        \[\frac{Essential + Worthwhile}{Essential + Worthwhile + Unimportant + Unwise}\]
    \item Percentage of '\textbf{Unwise}' responses among all the responses:
        \[\frac{Unwise}{Essential + Worthwhile + Unimportant + Unwise}\]
\end{itemize}

We rank each question based on the above percentages, with the top rank (\#1) having the highest percentage in a dimension (Essential, Worthwhile+, or Unwise).
Table \ref{tab:essential_ranked} and Table \ref{tab:unwise_ranked} presents the most desired (Top 10 Essential, Top 10 Worthwhile+) and the most undesired (Top 10 Unwise) descriptive questions. For all 171 questions and their rank, see the appendix.

\subsubsection{Rating by Demographics}
Unlike the Microsoft study, we did not have employee database to rank responses based on demographics, and privacy regulations prevented us from asking people-related aspects such as years of experience (another deviation from the original study).
Nonetheless, in both the initial and the rating survey, we asked the following professional background data from the participants:

\begin{itemize}
    \item \textit{Discipline:} Participants were asked to indicate their primary working area: \textit{Development}, \textit{Test}, \textit{Project Management}, \textit{Other Engineer} (e.g. architect, lead), or \textit{Other Non-Engineer} (only one selection was possible).  
    \item \textit{Current Role:} Participants were asked to indicate their current role: \textit{Individual Contributor}, \textit{Lead}, \textit{Architect}, \textit{Manager}, \textit{Executive}, or \textit{Other} (more selections were possible).
\end{itemize}

To investigate the relations of descriptive questions to professional background (discipline or current role), we built stepwise logistic regression models. 
We build our own models since the referenced study did not share scripts to run statistical tests although we did follow their procedure as is.
Stepwise regression eliminated professional backgrounds that did not improve the model for a given question and a response.
In addition, we removed professional backgrounds for which the coefficient in the model was not statistically significant at p-value < 0.01.
For each of the 171 questions, we built a model with Essential response (yes/no) as a dependent variable and professional background as independent variable. 
We built similar models for Worthwhile+ and Unwise responses. 
In total, we built 513 models, three for each of the 171 descriptive questions.

\begin{table*}[ht] \centering
\caption{\ING categories and questions mapped on to the 12 Microsoft categories}
\label{tab:categories}
\footnotesize
\begin{tabular}{llrr|r|r|rrrr|r}
& \multicolumn{5}{c|}{{\color[HTML]{333333} \textbf{\ING 2019 Study}}}                                           & \multicolumn{4}{c}{{\color[HTML]{333333} \textbf{Microsoft 2014 Study}}}
& \\ \hline
Category & & & Cards & 
\begin{tabular}[c]{@{}r@{}}Subcategories\end{tabular} & 
\begin{tabular}[c]{@{}r@{}}Descriptive\\ Questions\end{tabular} & & \multicolumn{1}{r|}{Cards} & \multicolumn{1}{r|}{\begin{tabular}[c]{@{}r@{}}Subcategories\end{tabular}} & \begin{tabular}[c]{@{}r@{}}Descriptive\\ Questions\end{tabular} & \begin{tabular}[c]{@{}r@{}}Difference \ING 2019 compared\\ to MS 2014\end{tabular} \\ \hline
Teams and Collaboration & TC & 14 & 4\% & 5 & 7 & 73 & \multicolumn{1}{r|}{10\%} & \multicolumn{1}{r|}{7} & 11 & $\downarrow$ 6\% \\
Testing Practices & TP & 32 & 9\% & 3 & 15 & 101 & \multicolumn{1}{r|}{14\%} & \multicolumn{1}{r|}{5} & 20 & $\downarrow$ 5\% \\
Services & SVC & 3 & 1\% & 2 & 1 & 42 & \multicolumn{1}{r|}{6\%} & \multicolumn{1}{r|}{2} & 8 & $\downarrow$ 5\% \\
Reuse and Shared Components & RSC & 5 & 1\% & 3 & 2 & 31 & \multicolumn{1}{r|}{4\%} & \multicolumn{1}{r|}{1} & 3 & $\downarrow$ 3\% \\ 
Customers and Requirements & CR & 9 & 3\% & 2 & 7 & 44 & \multicolumn{1}{r|}{6\%} & \multicolumn{1}{r|}{5} & 9 & $\downarrow$ 3\% \\
Software Development Lifecycle & SL & 6 & 2\% & 4 & 4 & 32 & \multicolumn{1}{r|}{4\%} & \multicolumn{1}{r|}{3} & 7 & $\downarrow$ 2\% \\
Development Practices & DP & 51 & 15\% & 14 & 38 & 117 & \multicolumn{1}{r|}{16\%} & \multicolumn{1}{r|}{13} & 28 & $\downarrow$ 1\% \\
Bug Measurements & BUG & 6 & 2\% & 3 & 5 & 23 & \multicolumn{1}{r|}{3\%} & \multicolumn{1}{r|}{4} & 7 & $\downarrow$ 1\% \\
Productivity & PROD & 29 & 9\% & 8 & 20 & 57 & \multicolumn{1}{r|}{8\%} & \multicolumn{1}{r|}{5} & 13 & $\uparrow$ 1\% \\
Evaluating Quality & EQ & 38 & 11\% & 6 & 11 & 47 & \multicolumn{1}{r|}{6\%} & \multicolumn{1}{r|}{6} & 16 & $\uparrow$ 5\% \\
Development Best Practices & BEST & 49 & 15\% & 7 & 36 & 65 & \multicolumn{1}{r|}{9\%} & \multicolumn{1}{r|}{6} & 9 & $\uparrow$ 6\% \\
Software Development Process & PROC & 46 & 14\% & 7 & 25 & 47 & \multicolumn{1}{r|}{6\%} & \multicolumn{1}{r|}{3} & 14 & $\uparrow$ 8\% \\ 
\hline
\textit{Discarded Cards} & & \textit{49} & \textit{15\%} & & & \textit{49} & \multicolumn{1}{r|}{\textit{7\%}} & \multicolumn{1}{r|}{\textit{}} & \textit{} & $\uparrow$ 8\% \\ 
\hline
Total Cards & & 337 & 100\% & 64 & 171 & 728 & \multicolumn{1}{r|}{100\%} & \multicolumn{1}{r|}{60} & 145 &
\end{tabular}
\begin{flushleft}
\footnotesize\emph{}Table sorted on the percentage difference in the number of questions in the \ING study compared to the Microsoft study.
\end{flushleft}
\end{table*}
\subsection{Comparison of Questions}
As a preliminary analysis, we start by looking at the similarities and differences in the broader themes or categories in both the studies.
Then for each theme, we see how the prominent questions in \ING compare against the prominent questions at Microsoft.

\begin{figure} [t] \centering
    \includegraphics[width=0.48\textwidth]
    {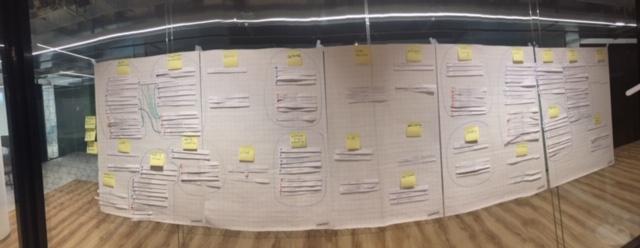}
  \caption{Analysis of \ING 2019 and MS 2014 questions.}
      \label{fig:Question_analysis}
\end{figure}

To make the comparison systematic, we followed a two-step approach.
First, we ran word counts on the questions from both the companies presenting a text-based comparison to identify broad differences.
Further, the first two authors manually analyzed top 100 essential questions from the two companies in detail.
The authors drew affinity diagrams using Microsoft questions (see Figure~\ref{fig:Question_analysis}) and appended related questions from \ING to it.
In case no cluster fits a question, a new cluster is created. 
This resulted in three types of clusters: match and no match (addition of \ING questions and deletion of Microsoft questions).
Analyses of the three clusters and the frequency distribution of questions (in addition to the previous three analyses) present insights into our research question. 

\section{Results}
\label{sec:results}
The original Microsoft study came up with 145 questions that software engineers want data scientists to answer.
Replicating the original study at \ING, we identified 171 data science questions.  

This section presents a comparison of the two sets of questions based on category, type of questions within categories, top-rated questions, bottom-rated questions, and questions relevant for different demographics.  
Next, we compare the questions from the two companies using word count and affinity diagrams to answer our research question. 

\subsection{Categories}
We noticed that some of our categories directly match the Microsoft study.
Other categories, however, can be mapped to one or more categories of the Microsoft study. 
No new emergent category in our study indicates that broadly there are no differences between the questions for a software-defined enterprise from a software company. 
For further analysis, we map our categories on to theirs, details on which are available in Table~\ref{tab:categories}.

Next, we explore the essential questions at \ING and their distinguishing link to the questions from the Microsoft study.
\vspace{2mm} 
\subsubsection{Bug Measures (BUG)}
The essential questions at \ING relate to the effort spent on bugs, methods to prevent security-related vulnerabilities, and the relationship between bugs and specific \ING-related development platforms.

\vspace{0.5mm} \noindent \faCommentO \textit{"How does the effort spent on fixing vulnerabilities and bugs relate to effort spent on writing software correctly from the start?"} 

\vspace{0.5mm} \noindent \faCommentO \textit{"What methods are most effective in preventing security-related vulnerabilities or bugs from being introduced in software code?"}


\vspace{2mm} 
\subsubsection{Development Practices (DP)}
The performance and productivity of DevOps teams was found in a number of questions including team happiness and work pleasure (\# 1 question), ways of decision making, non-overlapping development activities in the same environment,  product ownership and business responsibilities, licensing of tools, and the choice of a data modeling approach.

\vspace{0.5mm} \noindent \faCommentO \textit{"What factors affect the performance and productivity of DevOps teams with regard to team happiness and pleasure in your work?"}

\vspace{0.5mm} \noindent \faCommentO \textit{"What factors affect the performance and productivity of DevOps teams with regard to evidence-based decision-making versus decision-making based on expert opinions?"}

\vspace{0.5mm} \noindent \faCommentO \textit{"What factors affect the performance and productivity of DevOps teams with regard to simultaneous slow and fast developments at the same time in the same environment?"}


\vspace{2mm} 
\subsubsection{Development Best Practices (BEST)}
This category emphasized best (or worst) development practices relating to technology selection, effectiveness, and choice of tools.

\vspace{0.5mm} \noindent \faCommentO \textit{"How can we make sure that we build for re-usability and scalability?"}

\vspace{0.5mm} \noindent \faCommentO \textit{"What factors affect high performance teams?"}

\vspace{0.5mm} \noindent \faCommentO \textit{"When do you remove an old module that you think is not being used anymore?"}

\vspace{2mm} 
\subsubsection{Testing Practices (TP)}
Questions here ranged from automated test data generation, on-demand provisioning of test environments, testing of high volumes, to question like "should we let loose Chaos Monkey" \citep{Netflix2011} \citep{Netflix2019} 

\vspace{0.5mm} \noindent \faCommentO \textit{"To what extent does on-demand provisioning of development and test environments, including up-to-date data affect delivery of software solutions?"}

\vspace{0.5mm} \noindent \faCommentO \textit{"What factors affect performance testing on high data volumes?"}

\vspace{0.5mm} \noindent \faCommentO \textit{"How can a system for (semi) automated CRUD test data generation improve delivery of software solutions?"}

\vspace{0.5mm} \noindent \faCommentO \textit{"Should we let loose Chaos Monkey, like Netflix?} 

\vspace{2mm} 
\subsubsection{Evaluating Quality (EQ)}
This category included questions on code analysis, ways to assess quality of software code, and effectiveness of testing practices.  

\vspace{0.5mm} \noindent \faCommentO \textit{"What methods can be applied to analyze whether software code is working as expected?"}

\vspace{0.5mm} \noindent \faCommentO \textit{"To what extent does testability of software code affect the quality of code?"}




\vspace{2mm} 
\subsubsection{Customers and Requirements (CR)}
The essential questions related to measure customer value, requirement validation, and the use of formal models. Notably, questions relating to development trade-offs such as backward compatibility or the impact of testing in production appeared in the Microsoft study but not ours.

\vspace{0.5mm} \noindent \faCommentO \textit{"How to measure the customer value of a software product?"}

\vspace{0.5mm} \noindent \faCommentO \textit{"How can requirements be validated before starting actual software development?"}

\vspace{0.5mm} \noindent \faCommentO \textit{"How can user feedback be integrated in an efficient and effective way into software code?"}

\vspace{2mm} 
\subsubsection{Software Development Lifecycle (SL)}
Questions in this category related to the effectiveness and performance in lead time, cost, and quality (same as  the Microsoft study) but also questions relating to security and risk from a management perspective.

\vspace{0.5mm} \noindent \faCommentO \textit{"What factors affect providing new technologies to consumers, and can implementations of new technology be internally and externally benchmarked?"}

\vspace{0.5mm} \noindent \faCommentO \textit{"What factors affect estimation of lead time, cost, and quality of software deliveries?"}


\vspace{2mm} 
\subsubsection{Software Development Process (PROC)}
Our questions related to development processes, technology selection, and deployment of software solutions. At Microsoft, in contrast, questions related to the choice of software methodology (e.g. ways in which agile is better than waterfall? and benefits of pair programming). We also noticed that at \ING topics like the effects of automated continuous delivery pipeline popped up which were not seen in the Microsoft study.


\vspace{0.5mm} \noindent \faCommentO \textit{"How can we improve the deployment process in DevOps teams?"}

\vspace{0.5mm} \noindent \faCommentO \textit{"Does a focus on quick release of features and bug fixes into production help to achieve confidence and agility?"}



\vspace{2mm} 
\subsubsection{Productivity (PROD)}
 This category had questions on the productivity of DevOps teams - but also individual developers, ranked essential. Notably, questions related to the measurement of individual developers (e.g. the questions mentioned below regarding "great coder" and "open spaces") were often ranked "Unwise". 
Quite unlike the Microsoft study, where respondents considered these questions as unwise, engineers at \ING had a mixed opinion. 

\vspace{0.5mm} \noindent \faCommentO \textit{"What factors affect the performance of DevOps teams and the quality of software code with regard to quantity and quality of environments?"}

\vspace{0.5mm} \noindent \faCommentO \textit{"Are developers working in an open space with several teams more effective or less than developers working in a room with just their team?"}

\vspace{0.5mm} \noindent \faCommentO \textit{"What makes a great coder? What aspects affect the performance of DevOps teams and the quality of software with regard to characteristics of an individual software engineer?"}


\vspace{2mm} 
\subsubsection{Teams and Collaborations (TC)}
Essential questions here are typically about dependencies between teams, team composition, team maturity, and knowledge sharing among teams. 

\vspace{0.5mm} \noindent \faCommentO \textit{"To what extent do dependencies on other teams affect team performance?"}

\vspace{0.5mm} \noindent \faCommentO \textit{"How does team maturity affect code quality and incidents?"}

\vspace{0.5mm} \noindent \faCommentO \textit{"What factors affect the composition of DevOps teams?"}





\newcolumntype{P}[1]{>{\centering\arraybackslash}p{#1}} 
\begin{table*}[t]
\caption{Questions with the highest "Essential" and "Worthwhile or higher" percentages.}
\label{tab:essential_ranked}
\footnotesize
\begin{tabular}{p{0.01cm}p{0.4cm}p{7.4cm}p{1.0cm}|P{0.8cm}P{1.2cm}P{0.7cm}|P{0.7cm}P{1.2cm}P{0.8cm}}
\toprule
 & &  &  & \multicolumn{3}{c}{Percentages} & \multicolumn{3}{c}{Rank} \\ & & Question & Category & Essential & Worthwhile+ & Unwise & Essential & Worthwhile+ & Unwise \\ \midrule
\MineSign & Q143 & What factors affect the   performance and productivity of DevOps teams with regard to team happiness and pleasure in your work? & DP & 68.4\% & 94.7\% & 0.0\% & 1 & 9 & 68 \\
\faStar & Q98 & How does on-demand provisoning of develop- and test environments, including up-to-date data affect delivery of software solutions? & TP & 66,7\% & 77,8\% & 0,0\% & 2 & 95 & 68 \\
\faStar & Q37 & How can we make sure that we build for reusability and scalability? & BEST & 63.2\% & 89.5\% & 5.3\% & 3 & 42 & 63 \\
\MineSign & Q145 & What factors affect the performance of DevOps teams and the quality of software code with regard to quantity and quality of environments? & PROD & 60.0\% & 100.0\% & 0.0\% & 4 & 1 & 68 \\
\MineSign & Q114 & What factors affect High Performance Teams? & BEST & 58.8\% & 82.4\% & 0.0\% & 5 & 75 & 68 \\
\faStar & Q154 & What factors affect understandability and readability of software code for other developers? & DP & 58.3\% & 91.7\% & 8.3\% & 6 & 25 & 44 \\
\MineSign & Q76 & How can we improve the deployment process in DevOps teams? & PROC & 56.3\% & 93.8\% & 0.0\% & 7 & 15 & 68 \\
\faStar & Q36 & How does the effort spend on fixing vulnerabilities and bugs relate to effort spend on writing software correctly from the start? & BUG & 56.3\% & 93.8\% & 0.0\% & 7 & 15 & 68 \\
\MineSign & Q53 & How does a continuous delivery pipeline with automated testing and migrating, including rollback facilities affect the performance of DevOps teams and the quality of software? & PROC & 56.3\% & 93.8\% & 0.0\% & 7 & 15 & 68 \\
\faStar & Q22 & How can requirements be validated before starting actual software development? & CR & 55.6\% & 88.9\% & 0.0\% & 10 & 44 & 68 \\
\faStar & Q123 & What factors affect performance testing on high data volumes? & TP & 55.6\% & 88.9\% & 0.0\% & 10 & 44 & 68 \\ 
\faUser & Q58 & How to measure the customer value of a software product? & CR & 55.6\% & 77.8\% & 11.1\% & 10 & 95 & 20 \\
\midrule
\faStar & Q88 & To what extent does testability affect the quality of software code? & EQ & 52.9\% & 100.0\% & 0.0\% & 14 & 1 & 68 \\
\MineSign & Q67 & To what extent do automated checks of coding conventions, code quality, code complexity, and test-coverage affect the quality of software systems and the performance of DevOps teams? & EQ & 47.1\% & 100.0\% & 0.0\% & 25 & 1 & 68 \\
\faStar & Q11 & How can a system for (semi) automated CRUD test data generation improve delivery of software solutions? & TP & 44.4\% & 100.0\% & 0.0\% & 32 & 1 & 68 \\
\MineSign & Q104 & What aspects affect the performance of DevOps teams and the quality of software with regard to software architecture? & PROD & 40.0\% & 100.0\% & 0.0\% & 44 & 1 & 68 \\
\faStar & Q19 & How can editors help software developers to document their public functions in a way that it is available for other developers? & CR & 33.3\% & 100.0\% & 0.0\% & 73 & 1 & 68 \\
\faStar & Q122 & What factors affect maintainability of software systems? & EQ & 33.3\% & 100.0\% & 0.0\% & 73 & 1 & 68 \\
\faStar & Q80 & How do automated controls within the continuous delivery pipeline affect the effort spent on risk and security? & DP & 50.0\% & 95.0\% & 0.0\% & 15 & 8 & 68 \\
\bottomrule
\\[-2pt]
\end{tabular}
\begin{flushleft}
\footnotesize\emph{}Table is sorted on 
Rank Essential. The icon \faUser \,indicates customer related questions, \MineSign \,indicates questions that focus on the engineer and the effects of software development practices and processes on her work, and \faStar \,indicates quality related questions.
\end{flushleft}
\end{table*}

\subsection{Top-Rated Questions}
Table \ref{tab:essential_ranked} shows top 15  "Essential"  and top 10  "Worthwhile or higher" questions. 
Interestingly, only two out of the top 15 ``Essential'' questions were a part of the top 10 ``Worthwhile or higher'' questions and none vice-versa. 
This potentially means that our participants are more pronounced and opt for Essential or Worthwhile only when they feel so.
Culture can be another possible reason since all participants at \ING are located in one country while participants of the Microsoft study were more diverse~\citep{doi:10.1177/0022002184015004003}.

Our top questions are on development processes, technology selection, and deployment of software solutions.  
The top related questions at Microsoft, in contrast, relates to the choice of software methodology (e.g. ways in which agile is better than waterfall? and benefits of pair programming). 
We also noticed that in our study topics like the effects of automated continuous delivery pipeline popped up which were not seen in the Microsoft study.


Notably, a large fraction of the top 20 "Essential" or "Worthwhile or higher" questions at Microsoft (9 out of 20; including top 2) relates to customers. 
This suggests that for Microsoft customer benefit is most important or perhaps one of the most important question.
Our study, in contrast, paints a very different picture. 
Only two out of the 336 questions in the initial survey mentioned the word "customer" and only one of those questions made it to the top-20 (Q58 "How to measure the customer value of a software product" at rank 10 "Essential"). This question is, in line with the Microsoft study, marked with icon \faUser, in Table \ref{tab:essential_ranked}.

Another eight "Essential" or "Worthwhile or higher" questions in the Microsoft study (marked with icon \MineSign) focus on the engineer and the effects of software development practices and processes on her work. 
In our study, we identified nine questions with this icon. 
In addition to the focus on individual engineer, many of the questions in our study relates to the concept of the DevOps team. 
Overall, it seems that Microsoft has a big focus on customer while \ING emphasizes on the engineering team itself.
Finally, seven questions in the Microsoft study (marked with the icon \faStar) were about quality-related issues (same as ours with eleven questions).

\begin{table*}[t]
\caption{Questions with the highest "Unwise" percentages (opposition).}
\label{tab:unwise_ranked}
\footnotesize
\begin{tabular}{p{0.4cm}p{7.8cm}p{1.0cm}|P{0.8cm}P{1.2cm}p{0.7cm}|P{0.8cm}P{1.2cm}P{0.7cm}}
\toprule
 &  &  & \multicolumn{3}{c}{Percentages} & \multicolumn{3}{c}{Rank} \\ & Question & Category & Essential & Worthwhile+ & Unwise & Essential & Worthwhile+ & Unwise \\ \midrule
Q27 & How can software solutions in one common language be developed in a way that it is applicable to every person, regardless of ones interest in software development? & CR & 22.2\% & 55.6\% & 33.3\% & 121 & 152 & 1 \\
Q39 & How can Windows-server images be created in order to facilitate testing within a continuous delivery pipeline? & DP & 9.1\% & 45.5\% & 27.3\% & 162 & 163 & 2 \\
Q170 & Why do many developers focus on the newest of the newest? Why don't they leave this to a small group in order to use time and effort more efficiently? & DP & 21.1\% & 47.4\% & 26.3\% & 128 & 161 & 3 \\
Q161 & What makes a great coder? What aspects affect the performance of DevOps teams and the quality of software with regard to characteristics of an individual software engineer? & PROD & 44.4\% & 66.7\% & 22.2\% & 32 & 128 & 4 \\
Q134 & What factors affect TFS (Team Foundation Services) - a Microsoft product that provides source code management - with regard to working with automated pipelines? & BEST & 38.9\% & 72.2\% & 22.2\% & 54 & 118 & 4 \\
Q30 & How can the performance of individual software engineers be benchmarked internally \ING and externally with other companies? & PROD & 40.0\% & 50.0\% & 20.0\% & 44 & 157 & 6 \\
Q77 & To what extent does changing of requirements during development affect the delivery of software solutions? & PROC & 12.5\% & 68.8\% & 18.8\% & 150 & 124 & 7 \\
Q21 & How can PL1 software code be converted to Cobol code, while maintaining readability of the code in order to simplify an application environment? & BEST & 18.2\% & 36.4\% & 18.2\% & 140 & 169 & 8 \\
Q38 & How can we measure the time to market of software solutions delivered within a department at \ING in order to benchmark the performance of that department against others. & DP & 9.1\% & 54.5\% & 18.2\% & 162 & 155 & 8 \\
Q149 & What factors affect the use of machine learning in software development over a period of ten years? & DP & 16.7\% & 66.7\% & 16.7\% & 143 & 128 & 10 \\
Q28 & How can the cost of data be identified, in order to sign a price tag to data? & DP & 5.6\% & 50.0\% & 16.7\% & 168 & 157 & 10 \\ 
\bottomrule
\\[-2pt]
\end{tabular}
\begin{flushleft}
\footnotesize\emph{}Table is sorted on Rank Unwise.
\end{flushleft}
\end{table*}

\begin{table*}[ht]
\caption{Statistically significant rating differences for the response "Essential" by professional background.}
\label{tab:demographics_discipline}
\footnotesize
\begin{tabular}{p{0.4cm}p{9.2cm}p{1.2cm}|p{1.5cm}|P{1.0cm}P{1.1cm}P{0.9cm}}
\toprule
 &  &  &  & \multicolumn{3}{c}{Discipline} \\
 & Question & Category & Response & Dev & Test & PM \\
 \midrule
Q2 & Are there practices of good software teams from the perspective of releasing software solutions into production? & PROC & Essential & \textbf{66.7\%} & 5.6 \% & 11.1\% \\
Q21 & How can PL1 software code be converted to Cobol code, while maintaining readability of the code in order to simplify an application environment? & BEST & Essential & \textbf{66.7\%} & 4.8 \% & 0.0\% \\
Q28 & How can the cost of data be identified, in order to sign a price tag to data? & DP & Essential & \textbf{72.7\%} & 0.0 \% & 0.0\% \\
Q46 & How do static code analysis tools such as Fortify and Sonar influence the quality of software engineering products? & BEST & Essential & \textbf{36.6\%} & 0.0 \% & 27.3\% \\
Q88 & \textit{To what extent does testability affect the quality of software code?} & EQ & Essential & \textbf{68.4\%} & 0.0 \% & 0.0\% \\
Q89 & How does time spent - in terms of full-time versus part-time - of a Scrum master affect the delivery of software solutions? & PROC & Essential & \textbf{66.7\%} & 5.6 \% & 11.1\% \\
Q95 & To what extent do dependencies on other teams affect team performance? & TC & Essential & \textbf{68.4\%} & 0.0 \% & 0.0\% \\
Q97 & How does documentation during software maintenance affect delivery of software solutions? & TP & Essential & \textbf{50.0\%} & 0.0 \% & 0.0\% \\
Q110 & What factors affect data analytics with regard to the use of external sources - such as market research reports and follow market trends - and let individual teams handle their local evolution? & PROC & Essential & \textbf{66.7\%} & 5.6 \% & 11.1\% \\
Q162 & What methods are most effective in preventing security related vulnerabilities or bugs from being introduced in software code? & BUG & Essential & \textbf{68.4\%} & 0.0 \% & 0.0\% \\
\bottomrule
\\[-2pt]
\end{tabular}
\end{table*}

\begin{table*}[h]
\label{tab:demographics_role}
\footnotesize
\begin{tabular}{p{0.4cm}p{9.2cm}p{1.2cm}|p{1.5cm}|P{1.0cm}P{1.1cm}P{0.9cm}}
\toprule
 &  &  &  & \multicolumn{3}{c}{Current Role} \\
 & Question & Category & Response & Manager & Individual & Architect \\
 \midrule
Q2 & Are there practices of good software teams from the perspective of releasing software solutions into production? & PROC & Essential & 41.4\% & \textbf{44.8} \% & 6.9\% \\
Q46 & How do static code analysis tools such as Fortify and Sonar influence the quality of software engineering products? & BEST & Essential & \textbf{69.2\%} & 15.4 \% & 0.0\% \\
Q97 & How does documentation during software maintenance affect delivery of software solutions? & TP & Essential & 10.0\% & \textbf{60.0} \% & 20.0\% \\
Q153 & What factors affect trunk-based development - a source-control branching model, where developers collaborate on code in a single branch - with regard to quality of software code? & BEST & Essential & 22.6\% & \textbf{54.8} \% & 9.7\% \\
\bottomrule
\\[-2pt]
\end{tabular}
\begin{flushleft}
\footnotesize\emph{}The professional background with the highest rating is highlighted in \textbf{bold}. Questions that are also in Table \ref{tab:essential_ranked} are shown in \textit{italics}. The role "Manager" includes the responses for "Manager" and "Lead".
\end{flushleft}
\end{table*}

\subsection{Bottom-Rated Questions}
Table \ref{tab:unwise_ranked} shows the top 10 unwise questions.
The most "Unwise" question (Q27) at \ING is the use of domain-specific language for use by non-experts. 
In the Microsoft study, the top five "Unwise" questions were all about a fear that respondents had of being rated. 
This effect can be seen in our study too (two of the top ten unwise questions - Q161 and Q30 - relate to measuring the performance of individual engineers), but not nearly as strongly as in the Microsoft study. 
Respondents in our study are torn on this topic; Q161 and Q30 are ranked as "Unwise" by respectively 22.2\% and 20.0\% of the respondents, but also ranked as "Essential" by another group of 44.4\% and 40.0\% of the respondents.
Also, it was interesting to see that measuring and benchmarking time to market of software solutions (Q38) is one of the top 10 unwise questions. 
It indicates resistance against comparing departments based on key performance indicators like the time to market.

\subsection{Rating by Demographics}
Table \ref{tab:demographics_discipline} shows essential questions for different disciplines (Developer, Tester, Project Management) and roles (Manager, Individual Contributor, Architect).
The complete inventory of questions for "Worthwhile or higher" and "Unwise" responses is present in the appendix.

\subsubsection{Discipline}
Microsoft study showed tester as a specific discipline mainly interested in test suites, bugs, and product quality.
We do not see the discipline "tester" in our study. 
This can be seen in Table \ref{tab:demographics_discipline} in which overall scores relating to "Test" are low and highest for "Development". Software engineers in the DevOps teams at \ING consider themselves to be generic developers, and testing is an integrated part of the discipline "developer". Both developers and testers are for example significantly interested in the testability of software code, and the quality of software related to an agile way of working and working in DevOps teams. 
Other findings relate to developers being significantly interested in team performance, e.g. regarding practices of good software teams from the perspective of releasing software into production, the use of data analytics to improve individual teams, and dependencies on other teams. 

\subsubsection{Role}
More individual contributors (e.g. developers) than managers are interested in good practices for software teams to release software into production. 
More managers than individual contributors, on the other hand, are interested in how software can help realize new policies and changes in the way of working, the relationship between documentation and maintenance of software, and to what extent the use of static code analysis tools such as Fortify and Sonar can affect the quality of software.   

\subsection{Comparing \ING and Microsoft Questions}
A comparison of the top 15 words from each company (see Table~\ref{tab:wordcount}) shows that a majority of the popular themes are the same (e.g., code, test, software, and quality).
Subtle differences, however, exist relating to rank (words in italics do not make it to top-15 in another company) and to the usage of a word in the other company (underlined). 

A subset of these differences can be attributed to differences in terminology. 
For instance, Microsoft uses terms like employee/employees and team/teams, while their equivalents at \ING are team/squad and engineer. 
Apart from this, Microsoft questions focused more on bugs, cost, time, customers, and tools while \ING employees talked about version, problem, systems, process, and impact.

\begin{table}[b]
\caption{Top 15 words from questions at \ING and Microsoft}
\label{tab:wordcount}
\footnotesize
\begin{tabular}{p{2.4cm}P{1.0cm}|p{2.7cm}P{1.0cm}}
\toprule
\multicolumn{2}{c}{Microsoft 2014} & \multicolumn{2}{c}{\ING 2019} \\ 
Word & Count & Word & Count \\ \midrule
code / coding & 48 (19\%) & testing / debugging & 92 (14\%) \\
test / tests / testing & 39 (16\%) & code / coding & 87 (13\%) \\
software & 31 (13\%) & software & 76 (11\%) \\
\textit{employee / employees} & 16 (6\%) & team / squad & 72 (11\%) \\
quality & 13 (5\%) & development & 62 (9\%) \\
\textit{bugs} & 13 (5\%) &\textit {version / library} & 39 (6\%) \\
development & 12 (5\%) & \underline{data} & 37 (6\%) \\
\textit{cost} & 11 (4\%) & \underline{incident, issue}, \textit{problem} & 36 (5\%) \\
team / teams & 11 (4\%) & \underline{security} / risk & 34 (5\%) \\
\textit{time} & 10 (4\%) & \textit{system / systems} & 34 (5\%) \\
\textit{customer / customers} & 9 (4\%) & quality & 34 (5\%) \\
impact & 9 (4\%) & \underline{production} & 21 (3\%) \\
productivity & 9 (4\%) & \textit{engineer} & 14 (2\%) \\
project & 9 (4\%) & \textit{process} & 14 (2\%) \\
\textit{tools} & 7 (3\%) & \textit{impact} & 13 (2\%) \\
\bottomrule
\\[-2pt]
\end{tabular}
\begin{flushleft}
\footnotesize\emph{Top 15 words (sorted on count) from Microsoft 2014 and \ING 2019 study. Words in the top-15 of one company and not the other are printed in \textit{italic}. Words in one list and not the other are \underline{underlined}}.
\end{flushleft}
\end{table}

Next, we inferred 24 themes from the clusters in the affinity diagram organically merging into three broad categories: relating to code (like understanding code, testing, quality), developers (individual and team productivity) and customers (note that while customers did not make it to the top-10 essential questions, they were important in the top-100). 
The 24 themes are automated testing, testing, understanding code, documentation, formal methods, code review, debugging, risk, refactoring, deployment, bug fixing, legacy, software quality, requirement, release, cloud, customer, estimation, team productivity, employer productivity, cost, team awareness, and agile working.
Investigating each theme and category in detail, we noticed that despite minor differences in the individual questions (some questions are broad in one company and specific in another), largely the key questions remain the same. 
For instance, employees at both the companies find questions relating to team productivity and employee productivity important, and yet assessing and comparing individual employees is undesirable.
There were, however, subtle differences. 
For instance, in the Microsoft study, we noticed a few questions eliciting the need for agile (vs. waterfall) as well as automated testing.
In the \ING study, however, we do not see such questions.
Rather, we see questions relating to the functional aspects of agile and automated testing.
Another subtle difference between the two companies is relating to code size. 
While not stated explicitly, from the nature of questions, it seems that the software teams at Microsoft are dealing with a large legacy codebase. 
This was reflected in questions relating to team awareness, code monitoring, backward compatibility, and refactoring.
Such questions, however, did not occur in \ING.
Other than the above, we saw cloud-related questions appearing in the Microsoft study only, while deployment-related questions appeared in \ING only.
In a nutshell, the core software development challenges of \ING are consistent with Microsoft.
There are although some nuanced differences which relate to the evolution of software market in the last five years as well as differences in the characteristics of the two companies.

\section{Discussion}
In this section, we discuss potential explanations for the differences in the list of questions found in our study compared to the Microsoft study.
We saw questions eliciting the need of agile methods in the Microsoft study while at \ING the questions related to functional aspects.
Our hypothesis here is that in the last five years there has been a change in the market: while in 2014, the questions on the adoption of agile and automated testing were common, in 2019 agile and automated testing became the norm.

We noticed that many questions at Microsoft deal with the scale of legacy code while no such question appeared at \ING.
One potential explanation for the observation can be that software systems at \ING are not of the same scale as Microsoft. 
Nonetheless, it remains a lesson that in the next 10 years, \ING can also be dealing with the complexity of large code base as Microsoft is experiencing today.

Finally, some questions appeared in only one organization.
We believe that these observations have something to do with the individual practices followed at Microsoft and \ING.
The deployment-related questions at \ING might be a result of the adoption of continuous delivery as a service.
Surprisingly, we did not see any finance-related questions in the \ING study.
\ING is a finance-based company and we expected to see some issues relating to both finance and software appear.
We noticed that employees often talked about security, but no real finance-related questions appear.
One explanation for this observation can be that the data science challenges relating to software development are independent of the actual field to which it is applied.
Supporting this argument, 145 questions from Microsoft also did not bring up any product specific details. 
Another potential explanation can be that through our question we anchored our respondents into asking software development related questions only.

\subsection{Implications}

One of the key findings of this paper is a list of 171 questions that software engineers in a large, software-driven organization would like to see answered, in order to optimize their software development activities.
From this, we see implications both in terms of practice and industry.

From a practical perspective, our study offers a new way of thinking to software development organizations who care about their development processes.
The questions originally raised by Microsoft are not just relevant to one of the largest tech companies in the world, but also to large software-defined enterprises active outside the tech-sector proper.
Inspired by these questions, an organization may  select the most relevant ones, and seek ways to address them.
While some questions are fundamentally hard to answer, organizations can make a starting point by collecting relevant data about their development processes. This, then, can help to make the development process itself more and more data-driven.
This is exactly how \ING intends to use the questions, and we believe companies around the world can follow suit.

From a research perspective, we have seen that the original Microsoft study has generated a series of papers that apply some form of Machine Learning to address the challenges raised in that study.
In the research community, \emph{AI-for-Software-Engineering} is an increasingly important topic, with many papers appearing that seek to apply machine learning to address software engineering problems.
Our study aims to add urgency and direction to this emerging field, by highlighting not just which questions \emph{can} be answered, but which ones \emph{should} be answered, from a practitioner perspective.

\subsection{Threats to Validity}

While our study expands the external validity of the original study, the fact remains that the two lists of questions are based on just two companies, which are both large organizations with over 10,000 software developers.
Our study highlights relevance to the FinTech sector, but it would be interesting to see further replications, for example in the automotive or health care sector, with different regulatory and additional safety constraints.
We expect that many of the  questions are also relevant to smaller organizations, especially given the agile way of working at \ING. Nevertheless, it will be worthwhile to further explore this.

From a perspective of internal validity, creating codes independent of the prior study is challenging.
It is possible that the similarities and differences seen compared to the Microsoft study relates to factors (e.g. researcher bias) other than the actual data.
We tried mitigating it by limiting our exposure to the previous study, not involving authors from the Microsoft study, and multiple authors generating codes independently.
Nonetheless, these biases are likely to exist.

For reasons of replication, we have used where possible the same survey questions, method of analysis and division into work area and discipline as in the Microsoft study \citep{Begel:2014:ATQ:2568225.2568233}. Apart from positive effects, this choice also had a negative effect with regard to analysis of demographics, mainly due to the fact that \ING uses a different way of working, including corresponding roles and team structure, than within Microsoft. Especially mapping the professional background "Discipline" of the original study on the demographic "Discipline" as applied within \ING was challenging. 

\ING works with DevOps teams, where an engineer fulfills both the area of developer and that of tester. As a result, testers were under-represented in both of the surveys we conducted. As a mitigation measure we therefore opted for combining the results of developers and testers in the further analysis.

Another potential threat is sensitivity of the ranks which mostly occurs at the extreme sides of the ranking, when, e.g., none of the participants label a question as `Unwise'.
In our study, on average each question received $21,888/177=123$ ratings and hence sensitivity of ranks is unlikely.

The presented results are free from corporate influence including Microsoft. 
A number of stakeholders at \ING (CIO, Corporate Communications) reviewed the submitted paper and approved it without any changes. 
Nevertheless, self-censorship by the authors remains a potential threat. 
Furthermore, researchers may have their biases which can potentially influence the results.

As also emphasized in related work on replications \citep{Wohlin:2014:GSS:2601248.2601268} \citep{Shull2008} \citep{Juristo2012} \citep{Kitchenham2008} \citep{MILLER2005233} \citep{GOMEZ20141033}, our study seeks to replicate earlier findings in a different context (e.g. other companies) and during a different time (environments and perceptions of engineers do change over time). In order to facilitate future replication of our study, we make the total set of descriptive questions and additional info on results of our tests available in our technical report.

\section{Conclusion}
Conducted at \ING---a software-defined enterprise providing banking solutions---this study presents 171 questions that software engineers at \ING would like data scientists to answer.
This study is a replication of a similar study at software company Microsoft, which resulted in 145 questions for data scientists.
Further, we went a step beyond to investigate the applicability of Microsoft's questions in \ING, as well as changes in trends over the last five years.

We compared the two lists of questions and found that the core software development challenges (relating to code, developer, and customer) remain the same.
Nonetheless, we observed subtle differences relating to the technology and software process developments (e.g., currently the debate about agile versus waterfall is now largely absent) and differences in the two organizations (e.g., Microsoft's focus on solving problems with a large code bases and \ING's challenges with continuous deployment).
We complete our analysis with a report on the impact Microsoft 2014 study generated, also indicating the impact that our study is capable to generate.

A thorough understanding of key questions software engineers have that can be answered by data scientists is of crucial importance to both the research community and modern software engineering practice. Our study aims to contribute to this understanding. We call on other companies, large and small, to conduct a similar analysis, in order to transform a software engineering into a data-driven endeavour addressing the most pressing questions.

\section*{Acknowledgments}
The authors thank \ING and all participating engineers for sharing their experience and views with us. 
We thank the authors of the original Microsoft study for the permission to reuse their research design figure.

\balance
\bibliographystyle{ACM-Reference-Format}
\bibliography{bibliography}


\begin{thebibliography}{79}


\ifx \showCODEN    \undefined \def \showCODEN     #1{\unskip}     \fi
\ifx \showDOI      \undefined \def \showDOI       #1{#1}\fi
\ifx \showISBNx    \undefined \def \showISBNx     #1{\unskip}     \fi
\ifx \showISBNxiii \undefined \def \showISBNxiii  #1{\unskip}     \fi
\ifx \showISSN     \undefined \def \showISSN      #1{\unskip}     \fi
\ifx \showLCCN     \undefined \def \showLCCN      #1{\unskip}     \fi
\ifx \shownote     \undefined \def \shownote      #1{#1}          \fi
\ifx \showarticletitle \undefined \def \showarticletitle #1{#1}   \fi
\ifx \showURL      \undefined \def \showURL       {\relax}        \fi
\providecommand\bibfield[2]{#2}
\providecommand\bibinfo[2]{#2}
\providecommand\natexlab[1]{#1}
\providecommand\showeprint[2][]{arXiv:#2}

\bibitem[\protect\citeauthoryear{Amershi, Begel, Bird, DeLine, Gall, Kamar,
  Nagappan, Nushi, and Zimmermann}{Amershi et~al\mbox{.}}{2019}]%
        {amershi2019software}
\bibfield{author}{\bibinfo{person}{Saleema Amershi}, \bibinfo{person}{Andrew
  Begel}, \bibinfo{person}{Christian Bird}, \bibinfo{person}{Robert DeLine},
  \bibinfo{person}{Harald Gall}, \bibinfo{person}{Ece Kamar},
  \bibinfo{person}{Nachiappan Nagappan}, \bibinfo{person}{Besmira Nushi}, {and}
  \bibinfo{person}{Thomas Zimmermann}.} \bibinfo{year}{2019}\natexlab{}.
\newblock \showarticletitle{Software engineering for machine learning: A case
  study}. In \bibinfo{booktitle}{\emph{2019 IEEE/ACM 41st International
  Conference on Software Engineering: Software Engineering in Practice
  (ICSE-SEIP)}}. IEEE, \bibinfo{pages}{291--300}.
\newblock


\bibitem[\protect\citeauthoryear{Andrei, Calder, Chalmers, Morrison, and
  Rost}{Andrei et~al\mbox{.}}{2016}]%
        {andrei2016probabilistic}
\bibfield{author}{\bibinfo{person}{Oana Andrei}, \bibinfo{person}{Muffy
  Calder}, \bibinfo{person}{Matthew Chalmers}, \bibinfo{person}{Alistair
  Morrison}, {and} \bibinfo{person}{Mattias Rost}.}
  \bibinfo{year}{2016}\natexlab{}.
\newblock \showarticletitle{Probabilistic formal analysis of app usage to
  inform redesign}. In \bibinfo{booktitle}{\emph{International Conference on
  Integrated Formal Methods}}. Springer, \bibinfo{pages}{115--129}.
\newblock


\bibitem[\protect\citeauthoryear{Arndt}{Arndt}{2018}]%
        {arndt2018big}
\bibfield{author}{\bibinfo{person}{Timothy Arndt}.}
  \bibinfo{year}{2018}\natexlab{}.
\newblock \showarticletitle{Big Data and software engineering: prospects for
  mutual enrichment}.
\newblock \bibinfo{journal}{\emph{Iran Journal of Computer Science}}
  \bibinfo{volume}{1}, \bibinfo{number}{1} (\bibinfo{year}{2018}),
  \bibinfo{pages}{3--10}.
\newblock


\bibitem[\protect\citeauthoryear{Basiri, Hochstein, Jones, and Tucker}{Basiri
  et~al\mbox{.}}{2019}]%
        {Netflix2019}
\bibfield{author}{\bibinfo{person}{Ali Basiri}, \bibinfo{person}{Lorin
  Hochstein}, \bibinfo{person}{Nora Jones}, {and} \bibinfo{person}{Haley
  Tucker}.} \bibinfo{year}{2019}\natexlab{}.
\newblock \showarticletitle{Automating chaos experiments in production}.
\newblock \bibinfo{journal}{\emph{Proceedings of the 41st ACM/IEEE
  International Conference on Software Engineering (ICSE)}}
  (\bibinfo{year}{2019}).
\newblock


\bibitem[\protect\citeauthoryear{Batch and Elmqvist}{Batch and
  Elmqvist}{2017}]%
        {batch2017interactive}
\bibfield{author}{\bibinfo{person}{Andrea Batch} {and} \bibinfo{person}{Niklas
  Elmqvist}.} \bibinfo{year}{2017}\natexlab{}.
\newblock \showarticletitle{The interactive visualization gap in initial
  exploratory data analysis}.
\newblock \bibinfo{journal}{\emph{IEEE transactions on visualization and
  computer graphics}} \bibinfo{volume}{24}, \bibinfo{number}{1}
  (\bibinfo{year}{2017}), \bibinfo{pages}{278--287}.
\newblock


\bibitem[\protect\citeauthoryear{Begel and Zimmermann}{Begel and
  Zimmermann}{2014}]%
        {Begel:2014:ATQ:2568225.2568233}
\bibfield{author}{\bibinfo{person}{Andrew Begel} {and} \bibinfo{person}{Thomas
  Zimmermann}.} \bibinfo{year}{2014}\natexlab{}.
\newblock \showarticletitle{Analyze This! 145 Questions for Data Scientists in
  Software Engineering}. In \bibinfo{booktitle}{\emph{Proceedings of the 36th
  International Conference on Software Engineering}}
  \emph{(\bibinfo{series}{ICSE 2014})}. \bibinfo{publisher}{ACM},
  \bibinfo{address}{New York, NY, USA}, \bibinfo{pages}{12--23}.
\newblock
\showISBNx{978-1-4503-2756-5}
\urldef\tempurl%
\url{https://doi.org/10.1145/2568225.2568233}
\showDOI{\tempurl}


\bibitem[\protect\citeauthoryear{Beller, Gousios, Panichella, Proksch, Amann,
  and Zaidman}{Beller et~al\mbox{.}}{2017}]%
        {beller2017developer}
\bibfield{author}{\bibinfo{person}{Moritz Beller}, \bibinfo{person}{Georgios
  Gousios}, \bibinfo{person}{Annibale Panichella}, \bibinfo{person}{Sebastian
  Proksch}, \bibinfo{person}{Sven Amann}, {and} \bibinfo{person}{Andy
  Zaidman}.} \bibinfo{year}{2017}\natexlab{}.
\newblock \showarticletitle{Developer testing in the ide: Patterns, beliefs,
  and behavior}.
\newblock \bibinfo{journal}{\emph{IEEE Transactions on Software Engineering}}
  \bibinfo{volume}{45}, \bibinfo{number}{3} (\bibinfo{year}{2017}),
  \bibinfo{pages}{261--284}.
\newblock


\bibitem[\protect\citeauthoryear{Beller, Gousios, Panichella, and
  Zaidman}{Beller et~al\mbox{.}}{2015b}]%
        {beller2015and}
\bibfield{author}{\bibinfo{person}{Moritz Beller}, \bibinfo{person}{Georgios
  Gousios}, \bibinfo{person}{Annibale Panichella}, {and} \bibinfo{person}{Andy
  Zaidman}.} \bibinfo{year}{2015}\natexlab{b}.
\newblock \showarticletitle{When, how, and why developers (do not) test in
  their IDEs}. In \bibinfo{booktitle}{\emph{Proceedings of the 2015 10th Joint
  Meeting on Foundations of Software Engineering}}. ACM,
  \bibinfo{pages}{179--190}.
\newblock


\bibitem[\protect\citeauthoryear{Beller, Gousios, and Zaidman}{Beller
  et~al\mbox{.}}{2015a}]%
        {beller2015much}
\bibfield{author}{\bibinfo{person}{Moritz Beller}, \bibinfo{person}{Georgios
  Gousios}, {and} \bibinfo{person}{Andy Zaidman}.}
  \bibinfo{year}{2015}\natexlab{a}.
\newblock \showarticletitle{How (much) do developers test?}. In
  \bibinfo{booktitle}{\emph{2015 IEEE/ACM 37th IEEE International Conference on
  Software Engineering}}, Vol.~\bibinfo{volume}{2}. IEEE,
  \bibinfo{pages}{559--562}.
\newblock


\bibitem[\protect\citeauthoryear{Bird, Menzies, and Zimmermann}{Bird
  et~al\mbox{.}}{2015}]%
        {Bird20151}
\bibfield{author}{\bibinfo{person}{Christian Bird}, \bibinfo{person}{Tim
  Menzies}, {and} \bibinfo{person}{Thomas Zimmermann}.}
  \bibinfo{year}{2015}\natexlab{}.
\newblock \showarticletitle{Chapter 1 - Past, Present, and Future of Analyzing
  Software Data}.
\newblock In \bibinfo{booktitle}{\emph{The Art and Science of Analyzing
  Software Data}}, \bibfield{editor}{\bibinfo{person}{Christian Bird},
  \bibinfo{person}{Tim Menzies}, {and} \bibinfo{person}{Thomas Zimmermann}}
  (Eds.). \bibinfo{publisher}{Morgan Kaufmann}, \bibinfo{address}{Boston},
  \bibinfo{pages}{1 -- 13}.
\newblock
\showISBNx{978-0-12-411519-4}
\urldef\tempurl%
\url{https://doi.org/10.1016/B978-0-12-411519-4.00001-X}
\showDOI{\tempurl}


\bibitem[\protect\citeauthoryear{Cartaxo, Pinto, and Soares}{Cartaxo
  et~al\mbox{.}}{2018}]%
        {cartaxo2018role}
\bibfield{author}{\bibinfo{person}{Bruno Cartaxo}, \bibinfo{person}{Gustavo
  Pinto}, {and} \bibinfo{person}{Sergio Soares}.}
  \bibinfo{year}{2018}\natexlab{}.
\newblock \showarticletitle{The role of rapid reviews in supporting
  decision-making in software engineering practice}. In
  \bibinfo{booktitle}{\emph{Proceedings of the 22nd International Conference on
  Evaluation and Assessment in Software Engineering 2018}}.
  \bibinfo{pages}{24--34}.
\newblock


\bibitem[\protect\citeauthoryear{Carver, Dieste, Kraft, Lo, and
  Zimmermann}{Carver et~al\mbox{.}}{2016}]%
        {carver2016practitioners}
\bibfield{author}{\bibinfo{person}{Jeffrey~C Carver}, \bibinfo{person}{Oscar
  Dieste}, \bibinfo{person}{Nicholas~A Kraft}, \bibinfo{person}{David Lo},
  {and} \bibinfo{person}{Thomas Zimmermann}.} \bibinfo{year}{2016}\natexlab{}.
\newblock \showarticletitle{How practitioners perceive the relevance of esem
  research}. In \bibinfo{booktitle}{\emph{Proceedings of the 10th ACM/IEEE
  International Symposium on Empirical Software Engineering and Measurement}}.
  ACM, \bibinfo{pages}{56}.
\newblock


\bibitem[\protect\citeauthoryear{Chen, Fu, Krishna, and Menzies}{Chen
  et~al\mbox{.}}{2018}]%
        {chen2018applications}
\bibfield{author}{\bibinfo{person}{Di Chen}, \bibinfo{person}{Wei Fu},
  \bibinfo{person}{Rahul Krishna}, {and} \bibinfo{person}{Tim Menzies}.}
  \bibinfo{year}{2018}\natexlab{}.
\newblock \showarticletitle{Applications of psychological science for
  actionable analytics}. In \bibinfo{booktitle}{\emph{Proceedings of the 2018
  26th ACM Joint Meeting on European Software Engineering Conference and
  Symposium on the Foundations of Software Engineering}}.
  \bibinfo{pages}{456--467}.
\newblock


\bibitem[\protect\citeauthoryear{D{\k{a}}browski, Letier, Perini, and
  Susi}{D{\k{a}}browski et~al\mbox{.}}{2019}]%
        {dkabrowski2019finding}
\bibfield{author}{\bibinfo{person}{Jacek D{\k{a}}browski},
  \bibinfo{person}{Emmanuel Letier}, \bibinfo{person}{Anna Perini}, {and}
  \bibinfo{person}{Angelo Susi}.} \bibinfo{year}{2019}\natexlab{}.
\newblock \showarticletitle{Finding and analyzing app reviews related to
  specific features: A research preview}. In
  \bibinfo{booktitle}{\emph{International Working Conference on Requirements
  Engineering: Foundation for Software Quality}}. Springer,
  \bibinfo{pages}{183--189}.
\newblock


\bibitem[\protect\citeauthoryear{Deewattananon and Sammapun}{Deewattananon and
  Sammapun}{2017}]%
        {deewattananon2017analyzing}
\bibfield{author}{\bibinfo{person}{Boonyarit Deewattananon} {and}
  \bibinfo{person}{Usa Sammapun}.} \bibinfo{year}{2017}\natexlab{}.
\newblock \showarticletitle{Analyzing user reviews in Thai language toward
  aspects in mobile applications}. In \bibinfo{booktitle}{\emph{2017 14th
  International Joint Conference on Computer Science and Software Engineering
  (JCSSE)}}. IEEE, \bibinfo{pages}{1--6}.
\newblock


\bibitem[\protect\citeauthoryear{Denny, Becker, Craig, Wilson, and
  Banaszkiewicz}{Denny et~al\mbox{.}}{2019}]%
        {denny2019research}
\bibfield{author}{\bibinfo{person}{Paul Denny}, \bibinfo{person}{Brett~A
  Becker}, \bibinfo{person}{Michelle Craig}, \bibinfo{person}{Greg Wilson},
  {and} \bibinfo{person}{Piotr Banaszkiewicz}.}
  \bibinfo{year}{2019}\natexlab{}.
\newblock \showarticletitle{Research This! Questions That Computing Educators
  Most Want Computing Education Researchers to Answer}. In
  \bibinfo{booktitle}{\emph{Proceedings of the 2019 ACM Conference on
  International Computing Education Research}}. \bibinfo{pages}{259--267}.
\newblock


\bibitem[\protect\citeauthoryear{Devanbu, Zimmermann, and Bird}{Devanbu
  et~al\mbox{.}}{2016}]%
        {7886896}
\bibfield{author}{\bibinfo{person}{P. Devanbu}, \bibinfo{person}{T.
  Zimmermann}, {and} \bibinfo{person}{C. Bird}.}
  \bibinfo{year}{2016}\natexlab{}.
\newblock \showarticletitle{Belief Evidence in Empirical Software Engineering}.
  In \bibinfo{booktitle}{\emph{2016 IEEE/ACM 38th International Conference on
  Software Engineering (ICSE)}}. \bibinfo{pages}{108--119}.
\newblock
\urldef\tempurl%
\url{https://doi.org/10.1145/2884781.2884812}
\showDOI{\tempurl}


\bibitem[\protect\citeauthoryear{Ford and Parnin}{Ford and Parnin}{2015}]%
        {ford2015exploring}
\bibfield{author}{\bibinfo{person}{Denae Ford} {and} \bibinfo{person}{Chris
  Parnin}.} \bibinfo{year}{2015}\natexlab{}.
\newblock \showarticletitle{Exploring causes of frustration for software
  developers}. In \bibinfo{booktitle}{\emph{2015 IEEE/ACM 8th International
  Workshop on Cooperative and Human Aspects of Software Engineering}}. IEEE,
  \bibinfo{pages}{115--116}.
\newblock


\bibitem[\protect\citeauthoryear{Ford, Smith, Guo, and Parnin}{Ford
  et~al\mbox{.}}{2016}]%
        {ford2016paradise}
\bibfield{author}{\bibinfo{person}{Denae Ford}, \bibinfo{person}{Justin Smith},
  \bibinfo{person}{Philip~J Guo}, {and} \bibinfo{person}{Chris Parnin}.}
  \bibinfo{year}{2016}\natexlab{}.
\newblock \showarticletitle{Paradise unplugged: Identifying barriers for female
  participation on stack overflow}. In \bibinfo{booktitle}{\emph{Proceedings of
  the 2016 24th ACM SIGSOFT International Symposium on Foundations of Software
  Engineering}}. \bibinfo{pages}{846--857}.
\newblock


\bibitem[\protect\citeauthoryear{Fu and Menzies}{Fu and Menzies}{2017}]%
        {fu2017easy}
\bibfield{author}{\bibinfo{person}{Wei Fu} {and} \bibinfo{person}{Tim
  Menzies}.} \bibinfo{year}{2017}\natexlab{}.
\newblock \showarticletitle{Easy over hard: A case study on deep learning}. In
  \bibinfo{booktitle}{\emph{Proceedings of the 2017 11th joint meeting on
  foundations of software engineering}}. \bibinfo{pages}{49--60}.
\newblock


\bibitem[\protect\citeauthoryear{Garousi, Co{\c{s}}kun{\c{c}}ay, Betin-Can, and
  Demir{\"o}rs}{Garousi et~al\mbox{.}}{2015}]%
        {garousi2015survey}
\bibfield{author}{\bibinfo{person}{Vahid Garousi}, \bibinfo{person}{Ahmet
  Co{\c{s}}kun{\c{c}}ay}, \bibinfo{person}{Aysu Betin-Can}, {and}
  \bibinfo{person}{Onur Demir{\"o}rs}.} \bibinfo{year}{2015}\natexlab{}.
\newblock \showarticletitle{A survey of software engineering practices in
  Turkey}.
\newblock \bibinfo{journal}{\emph{Journal of Systems and Software}}
  \bibinfo{volume}{108} (\bibinfo{year}{2015}), \bibinfo{pages}{148--177}.
\newblock


\bibitem[\protect\citeauthoryear{Garousi and Felderer}{Garousi and
  Felderer}{2017}]%
        {garousi2017worlds}
\bibfield{author}{\bibinfo{person}{Vahid Garousi} {and}
  \bibinfo{person}{Michael Felderer}.} \bibinfo{year}{2017}\natexlab{}.
\newblock \showarticletitle{Worlds apart: industrial and academic focus areas
  in software testing}.
\newblock \bibinfo{journal}{\emph{IEEE Software}} \bibinfo{volume}{34},
  \bibinfo{number}{5} (\bibinfo{year}{2017}), \bibinfo{pages}{38--45}.
\newblock


\bibitem[\protect\citeauthoryear{Garousi and Herkiloglu}{Garousi and
  Herkiloglu}{2016}]%
        {garousi2016selecting}
\bibfield{author}{\bibinfo{person}{Vahid Garousi} {and} \bibinfo{person}{Kadir
  Herkiloglu}.} \bibinfo{year}{2016}\natexlab{}.
\newblock \showarticletitle{Selecting the right topics for industry-academia
  collaborations in software testing: an experience report}. In
  \bibinfo{booktitle}{\emph{2016 IEEE International Conference on Software
  Testing, Verification and Validation (ICST)}}. IEEE,
  \bibinfo{pages}{213--222}.
\newblock


\bibitem[\protect\citeauthoryear{German, Robles, Poo-Caama{\~n}o, Yang, Iida,
  and Inoue}{German et~al\mbox{.}}{2018}]%
        {german2018my}
\bibfield{author}{\bibinfo{person}{Daniel German}, \bibinfo{person}{Gregorio
  Robles}, \bibinfo{person}{Germ{\'a}n Poo-Caama{\~n}o}, \bibinfo{person}{Xin
  Yang}, \bibinfo{person}{Hajimu Iida}, {and} \bibinfo{person}{Katsuro Inoue}.}
  \bibinfo{year}{2018}\natexlab{}.
\newblock \showarticletitle{" Was My Contribution Fairly Reviewed?" A Framework
  to Study the Perception of Fairness in Modern Code Reviews}. In
  \bibinfo{booktitle}{\emph{2018 IEEE/ACM 40th International Conference on
  Software Engineering (ICSE)}}. IEEE, \bibinfo{pages}{523--534}.
\newblock


\bibitem[\protect\citeauthoryear{G\'{o}mez, Juristo, and Vegas}{G\'{o}mez
  et~al\mbox{.}}{2014}]%
        {GOMEZ20141033}
\bibfield{author}{\bibinfo{person}{Omar~S. G\'{o}mez}, \bibinfo{person}{Natalia
  Juristo}, {and} \bibinfo{person}{Sira Vegas}.}
  \bibinfo{year}{2014}\natexlab{}.
\newblock \showarticletitle{Understanding replication of experiments in
  software engineering: A classification}.
\newblock \bibinfo{journal}{\emph{Information and Software Technology}}
  \bibinfo{volume}{56}, \bibinfo{number}{8} (\bibinfo{year}{2014}),
  \bibinfo{pages}{1033 -- 1048}.
\newblock
\showISSN{0950-5849}
\urldef\tempurl%
\url{https://doi.org/10.1016/j.infsof.2014.04.004}
\showDOI{\tempurl}


\bibitem[\protect\citeauthoryear{Gousios, Safaric, and Visser}{Gousios
  et~al\mbox{.}}{2016}]%
        {gousios2016streaming}
\bibfield{author}{\bibinfo{person}{Georgios Gousios}, \bibinfo{person}{Dominik
  Safaric}, {and} \bibinfo{person}{Joost Visser}.}
  \bibinfo{year}{2016}\natexlab{}.
\newblock \showarticletitle{Streaming software analytics}. In
  \bibinfo{booktitle}{\emph{2016 IEEE/ACM 2nd International Workshop on Big
  Data Software Engineering (BIGDSE)}}. IEEE, \bibinfo{pages}{8--11}.
\newblock


\bibitem[\protect\citeauthoryear{Gu and Kim}{Gu and Kim}{2015}]%
        {gu2015parts}
\bibfield{author}{\bibinfo{person}{Xiaodong Gu} {and} \bibinfo{person}{Sunghun
  Kim}.} \bibinfo{year}{2015}\natexlab{}.
\newblock \showarticletitle{" What Parts of Your Apps are Loved by Users?"(T)}.
  In \bibinfo{booktitle}{\emph{2015 30th IEEE/ACM International Conference on
  Automated Software Engineering (ASE)}}. IEEE, \bibinfo{pages}{760--770}.
\newblock


\bibitem[\protect\citeauthoryear{Gupta, Sureka, Padmanabhuni, and
  Asadullah}{Gupta et~al\mbox{.}}{2015}]%
        {Gupta:2015:ISP:2820518.2820560}
\bibfield{author}{\bibinfo{person}{Monika Gupta}, \bibinfo{person}{Ashish
  Sureka}, \bibinfo{person}{Srinivas Padmanabhuni}, {and}
  \bibinfo{person}{Allahbaksh~Mohammedali Asadullah}.}
  \bibinfo{year}{2015}\natexlab{}.
\newblock \showarticletitle{Identifying Software Process Management Challenges:
  Survey of Practitioners in a Large Global IT Company}. In
  \bibinfo{booktitle}{\emph{Proceedings of the 12th Working Conference on
  Mining Software Repositories}} \emph{(\bibinfo{series}{MSR '15})}.
  \bibinfo{publisher}{IEEE Press}, \bibinfo{address}{Piscataway, NJ, USA},
  \bibinfo{pages}{346--356}.
\newblock
\showISBNx{978-0-7695-5594-2}


\bibitem[\protect\citeauthoryear{Guzman, El-Haliby, and Bruegge}{Guzman
  et~al\mbox{.}}{2015}]%
        {guzman2015ensemble}
\bibfield{author}{\bibinfo{person}{Emitza Guzman}, \bibinfo{person}{Muhammad
  El-Haliby}, {and} \bibinfo{person}{Bernd Bruegge}.}
  \bibinfo{year}{2015}\natexlab{}.
\newblock \showarticletitle{Ensemble methods for app review classification: An
  approach for software evolution (n)}. In \bibinfo{booktitle}{\emph{2015 30th
  IEEE/ACM International Conference on Automated Software Engineering (ASE)}}.
  IEEE, \bibinfo{pages}{771--776}.
\newblock


\bibitem[\protect\citeauthoryear{Hahn, Trapp, Wuttke, and D{\"o}llner}{Hahn
  et~al\mbox{.}}{2015}]%
        {hahn2015thread}
\bibfield{author}{\bibinfo{person}{Sebastian Hahn}, \bibinfo{person}{Matthias
  Trapp}, \bibinfo{person}{Nikolai Wuttke}, {and} \bibinfo{person}{J{\"u}rgen
  D{\"o}llner}.} \bibinfo{year}{2015}\natexlab{}.
\newblock \showarticletitle{Thread City: Combined Visualization of Structure
  and Activity for the Exploration of Multi-threaded Software Systems}. In
  \bibinfo{booktitle}{\emph{2015 19th International Conference on Information
  Visualisation}}. IEEE, \bibinfo{pages}{101--106}.
\newblock


\bibitem[\protect\citeauthoryear{Hassan and Xie}{Hassan and Xie}{2010}]%
        {hassan2010software}
\bibfield{author}{\bibinfo{person}{Ahmed~E Hassan} {and} \bibinfo{person}{Tao
  Xie}.} \bibinfo{year}{2010}\natexlab{}.
\newblock \showarticletitle{Software intelligence: the future of mining
  software engineering data}. In \bibinfo{booktitle}{\emph{Proceedings of the
  FSE/SDP workshop on Future of software engineering research}}. ACM,
  \bibinfo{pages}{161--166}.
\newblock


\bibitem[\protect\citeauthoryear{Hilton, Nelson, McDonald, McDonald, Metoyer,
  and Dig}{Hilton et~al\mbox{.}}{2016}]%
        {hilton2016tddviz}
\bibfield{author}{\bibinfo{person}{Michael Hilton}, \bibinfo{person}{Nicholas
  Nelson}, \bibinfo{person}{Hugh McDonald}, \bibinfo{person}{Sean McDonald},
  \bibinfo{person}{Ron Metoyer}, {and} \bibinfo{person}{Danny Dig}.}
  \bibinfo{year}{2016}\natexlab{}.
\newblock \showarticletitle{Tddviz: Using software changes to understand
  conformance to test driven development}. In
  \bibinfo{booktitle}{\emph{International Conference on Agile Software
  Development}}. Springer, Cham, \bibinfo{pages}{53--65}.
\newblock


\bibitem[\protect\citeauthoryear{Hofstede and Bond}{Hofstede and Bond}{1984}]%
        {doi:10.1177/0022002184015004003}
\bibfield{author}{\bibinfo{person}{Geert Hofstede} {and}
  \bibinfo{person}{Michael~H. Bond}.} \bibinfo{year}{1984}\natexlab{}.
\newblock \showarticletitle{Hofstede's Culture Dimensions: An Independent
  Validation Using Rokeach's Value Survey}.
\newblock \bibinfo{journal}{\emph{Journal of Cross-Cultural Psychology}}
  \bibinfo{volume}{15}, \bibinfo{number}{4} (\bibinfo{year}{1984}),
  \bibinfo{pages}{417--433}.
\newblock
\urldef\tempurl%
\url{https://doi.org/10.1177/0022002184015004003}
\showDOI{\tempurl}
\showeprint{https://doi.org/10.1177/0022002184015004003}


\bibitem[\protect\citeauthoryear{Izrailevsky and Tseitlin}{Izrailevsky and
  Tseitlin}{2011}]%
        {Netflix2011}
\bibfield{author}{\bibinfo{person}{Yury Izrailevsky} {and}
  \bibinfo{person}{Ariel Tseitlin}.} \bibinfo{year}{2011}\natexlab{}.
\newblock \showarticletitle{The Netflix Simian Army}.
\newblock \bibinfo{journal}{\emph{Netflix Technology Blog}}
  (\bibinfo{year}{2011}).
\newblock
\urldef\tempurl%
\url{https://medium.com/netflix-techblog/the-netflix-simian-army-16e57fbab116}
\showURL{%
\tempurl}


\bibitem[\protect\citeauthoryear{Juristo and G\'{o}mez}{Juristo and
  G\'{o}mez}{2012}]%
        {Juristo2012}
\bibfield{author}{\bibinfo{person}{Natalia Juristo} {and}
  \bibinfo{person}{Omar~S. G\'{o}mez}.} \bibinfo{year}{2012}\natexlab{}.
\newblock \bibinfo{booktitle}{\emph{Replication of Software Engineering
  Experiments}}.
\newblock \bibinfo{publisher}{Springer Berlin Heidelberg},
  \bibinfo{address}{Berlin, Heidelberg}, \bibinfo{pages}{60--88}.
\newblock
\showISBNx{978-3-642-25231-0}
\urldef\tempurl%
\url{https://doi.org/10.1007/978-3-642-25231-0-2}
\showDOI{\tempurl}


\bibitem[\protect\citeauthoryear{Kabeer, Nayebi, Ruhe, Carlson, and
  Chew}{Kabeer et~al\mbox{.}}{2017}]%
        {kabeer2017predicting}
\bibfield{author}{\bibinfo{person}{Shaikh~Jeeshan Kabeer},
  \bibinfo{person}{Maleknaz Nayebi}, \bibinfo{person}{Guenther Ruhe},
  \bibinfo{person}{Chris Carlson}, {and} \bibinfo{person}{Francis Chew}.}
  \bibinfo{year}{2017}\natexlab{}.
\newblock \showarticletitle{Predicting the vector impact of change-an
  industrial case study at brightsquid}. In \bibinfo{booktitle}{\emph{2017
  ACM/IEEE International Symposium on Empirical Software Engineering and
  Measurement (ESEM)}}. IEEE, \bibinfo{pages}{131--140}.
\newblock


\bibitem[\protect\citeauthoryear{K{\"a}fer}{K{\"a}fer}{2017}]%
        {kafer2017summarizing}
\bibfield{author}{\bibinfo{person}{Verena K{\"a}fer}.}
  \bibinfo{year}{2017}\natexlab{}.
\newblock \showarticletitle{Summarizing software engineering communication
  artifacts from different sources}. In \bibinfo{booktitle}{\emph{Proceedings
  of the 2017 11th Joint Meeting on Foundations of Software Engineering}}.
  \bibinfo{pages}{1038--1041}.
\newblock


\bibitem[\protect\citeauthoryear{Kano, Seraku, Takahashi, and ichi Tsuji}{Kano
  et~al\mbox{.}}{1984}]%
        {Kano1984AttractiveQA}
\bibfield{author}{\bibinfo{person}{Noriaki Kano}, \bibinfo{person}{Nobuhiko
  Seraku}, \bibinfo{person}{Fumio Takahashi}, {and} \bibinfo{person}{Shin ichi
  Tsuji}.} \bibinfo{year}{1984}\natexlab{}.
\newblock \showarticletitle{Attractive Quality and Must-Be Quality}.
\newblock \bibinfo{journal}{\emph{Journal of the Japanese Society for Quality
  Control}}  \bibinfo{volume}{14} (\bibinfo{year}{1984}),
  \bibinfo{pages}{39--48}.
\newblock


\bibitem[\protect\citeauthoryear{Khomh, Adams, Cheng, Fokaefs, and
  Antoniol}{Khomh et~al\mbox{.}}{2018}]%
        {khomh2018software}
\bibfield{author}{\bibinfo{person}{Foutse Khomh}, \bibinfo{person}{Bram Adams},
  \bibinfo{person}{Jinghui Cheng}, \bibinfo{person}{Marios Fokaefs}, {and}
  \bibinfo{person}{Giuliano Antoniol}.} \bibinfo{year}{2018}\natexlab{}.
\newblock \showarticletitle{Software engineering for machine-learning
  applications: The road ahead}.
\newblock \bibinfo{journal}{\emph{IEEE Software}} \bibinfo{volume}{35},
  \bibinfo{number}{5} (\bibinfo{year}{2018}), \bibinfo{pages}{81--84}.
\newblock


\bibitem[\protect\citeauthoryear{Kim, Zimmermann, DeLine, and Begel}{Kim
  et~al\mbox{.}}{2016a}]%
        {Kim:2016:ERD:2884781.2884783}
\bibfield{author}{\bibinfo{person}{Miryung Kim}, \bibinfo{person}{Thomas
  Zimmermann}, \bibinfo{person}{Robert DeLine}, {and} \bibinfo{person}{Andrew
  Begel}.} \bibinfo{year}{2016}\natexlab{a}.
\newblock \showarticletitle{The Emerging Role of Data Scientists on Software
  Development Teams}. In \bibinfo{booktitle}{\emph{Proceedings of the 38th
  International Conference on Software Engineering}}
  \emph{(\bibinfo{series}{ICSE '16})}. \bibinfo{publisher}{ACM},
  \bibinfo{address}{New York, NY, USA}, \bibinfo{pages}{96--107}.
\newblock
\showISBNx{978-1-4503-3900-1}
\urldef\tempurl%
\url{https://doi.org/10.1145/2884781.2884783}
\showDOI{\tempurl}


\bibitem[\protect\citeauthoryear{Kim, Zimmermann, DeLine, and Begel}{Kim
  et~al\mbox{.}}{2016b}]%
        {kim2016emerging}
\bibfield{author}{\bibinfo{person}{Miryung Kim}, \bibinfo{person}{Thomas
  Zimmermann}, \bibinfo{person}{Robert DeLine}, {and} \bibinfo{person}{Andrew
  Begel}.} \bibinfo{year}{2016}\natexlab{b}.
\newblock \showarticletitle{The emerging role of data scientists on software
  development teams}. In \bibinfo{booktitle}{\emph{Proceedings of the 38th
  International Conference on Software Engineering}}. ACM,
  \bibinfo{pages}{96--107}.
\newblock


\bibitem[\protect\citeauthoryear{Kitchenham}{Kitchenham}{2008}]%
        {Kitchenham2008}
\bibfield{author}{\bibinfo{person}{Barbara Kitchenham}.}
  \bibinfo{year}{2008}\natexlab{}.
\newblock \showarticletitle{The role of replications in empirical software
  engineering - a word of warning}.
\newblock \bibinfo{journal}{\emph{Empirical Software Engineering}}
  \bibinfo{volume}{13}, \bibinfo{number}{2} (\bibinfo{year}{2008}),
  \bibinfo{pages}{219--221}.
\newblock
\urldef\tempurl%
\url{https://doi.org/10.1007/s10664-008-9061-0}
\showDOI{\tempurl}


\bibitem[\protect\citeauthoryear{Kitchenham and Pfleeger}{Kitchenham and
  Pfleeger}{2008}]%
        {KitchenhamPfleeger2008}
\bibfield{author}{\bibinfo{person}{B. Kitchenham} {and} \bibinfo{person}{S.
  Pfleeger}.} \bibinfo{year}{2008}\natexlab{}.
\newblock \showarticletitle{Personal Opinion Surveys}.
\newblock \bibinfo{journal}{\emph{Guide to Advanced Empirical Software
  Engineering}} (\bibinfo{year}{2008}).
\newblock


\bibitem[\protect\citeauthoryear{Kochhar, Xia, Lo, and Li}{Kochhar
  et~al\mbox{.}}{2016}]%
        {kochhar2016practitioners}
\bibfield{author}{\bibinfo{person}{Pavneet~Singh Kochhar}, \bibinfo{person}{Xin
  Xia}, \bibinfo{person}{David Lo}, {and} \bibinfo{person}{Shanping Li}.}
  \bibinfo{year}{2016}\natexlab{}.
\newblock \showarticletitle{Practitioners' expectations on automated fault
  localization}. In \bibinfo{booktitle}{\emph{Proceedings of the 25th
  International Symposium on Software Testing and Analysis}}. ACM,
  \bibinfo{pages}{165--176}.
\newblock


\bibitem[\protect\citeauthoryear{Kononenko, Baysal, and Godfrey}{Kononenko
  et~al\mbox{.}}{2016}]%
        {kononenko2016code}
\bibfield{author}{\bibinfo{person}{Oleksii Kononenko}, \bibinfo{person}{Olga
  Baysal}, {and} \bibinfo{person}{Michael~W Godfrey}.}
  \bibinfo{year}{2016}\natexlab{}.
\newblock \showarticletitle{Code review quality: how developers see it}. In
  \bibinfo{booktitle}{\emph{2016 IEEE/ACM 38th International Conference on
  Software Engineering (ICSE)}}. IEEE, \bibinfo{pages}{1028--1038}.
\newblock


\bibitem[\protect\citeauthoryear{Krishna}{Krishna}{2017}]%
        {krishna2017learning}
\bibfield{author}{\bibinfo{person}{Rahul Krishna}.}
  \bibinfo{year}{2017}\natexlab{}.
\newblock \showarticletitle{Learning effective changes for software projects}.
  In \bibinfo{booktitle}{\emph{2017 32nd IEEE/ACM International Conference on
  Automated Software Engineering (ASE)}}. IEEE, \bibinfo{pages}{1002--1005}.
\newblock


\bibitem[\protect\citeauthoryear{Krishna and Menzies}{Krishna and
  Menzies}{2018}]%
        {krishna2018bellwethers}
\bibfield{author}{\bibinfo{person}{Rahul Krishna} {and} \bibinfo{person}{Tim
  Menzies}.} \bibinfo{year}{2018}\natexlab{}.
\newblock \showarticletitle{Bellwethers: A baseline method for transfer
  learning}.
\newblock \bibinfo{journal}{\emph{IEEE Transactions on Software Engineering}}
  \bibinfo{volume}{45}, \bibinfo{number}{11} (\bibinfo{year}{2018}),
  \bibinfo{pages}{1081--1105}.
\newblock


\bibitem[\protect\citeauthoryear{Leitner, Cito, and St{\"o}ckli}{Leitner
  et~al\mbox{.}}{2016}]%
        {leitner2016modelling}
\bibfield{author}{\bibinfo{person}{Philipp Leitner},
  \bibinfo{person}{J{\"u}rgen Cito}, {and} \bibinfo{person}{Emanuel
  St{\"o}ckli}.} \bibinfo{year}{2016}\natexlab{}.
\newblock \showarticletitle{Modelling and managing deployment costs of
  microservice-based cloud applications}. In
  \bibinfo{booktitle}{\emph{Proceedings of the 9th International Conference on
  Utility and Cloud Computing}}. ACM, \bibinfo{pages}{165--174}.
\newblock


\bibitem[\protect\citeauthoryear{Li, Ko, and Zhu}{Li et~al\mbox{.}}{2015}]%
        {li2015makes}
\bibfield{author}{\bibinfo{person}{Paul~Luo Li}, \bibinfo{person}{Andrew~J Ko},
  {and} \bibinfo{person}{Jiamin Zhu}.} \bibinfo{year}{2015}\natexlab{}.
\newblock \showarticletitle{What makes a great software engineer?}. In
  \bibinfo{booktitle}{\emph{Proceedings of the 37th International Conference on
  Software Engineering-Volume 1}}. IEEE Press, \bibinfo{pages}{700--710}.
\newblock


\bibitem[\protect\citeauthoryear{Lo, Nagappan, and Zimmermann}{Lo
  et~al\mbox{.}}{2015}]%
        {lo2015practitioners}
\bibfield{author}{\bibinfo{person}{David Lo}, \bibinfo{person}{Nachiappan
  Nagappan}, {and} \bibinfo{person}{Thomas Zimmermann}.}
  \bibinfo{year}{2015}\natexlab{}.
\newblock \showarticletitle{How practitioners perceive the relevance of
  software engineering research}. In \bibinfo{booktitle}{\emph{Proceedings of
  the 2015 10th Joint Meeting on Foundations of Software Engineering}}. ACM,
  \bibinfo{pages}{415--425}.
\newblock


\bibitem[\protect\citeauthoryear{Lopez-Martin, Chavoya, and
  Meda-Campa{\~n}a}{Lopez-Martin et~al\mbox{.}}{2014}]%
        {lopez2014machine}
\bibfield{author}{\bibinfo{person}{Cuauhtemoc Lopez-Martin},
  \bibinfo{person}{Arturo Chavoya}, {and} \bibinfo{person}{Maria~Elena
  Meda-Campa{\~n}a}.} \bibinfo{year}{2014}\natexlab{}.
\newblock \showarticletitle{A machine learning technique for predicting the
  productivity of practitioners from individually developed software projects}.
  In \bibinfo{booktitle}{\emph{15th IEEE/ACIS International Conference on
  Software Engineering, Artificial Intelligence, Networking and
  Parallel/Distributed Computing (SNPD)}}. IEEE, \bibinfo{pages}{1--6}.
\newblock


\bibitem[\protect\citeauthoryear{Lwakatare, Raj, Bosch, Olsson, and
  Crnkovic}{Lwakatare et~al\mbox{.}}{2019}]%
        {lwakatare2019taxonomy}
\bibfield{author}{\bibinfo{person}{Lucy~Ellen Lwakatare},
  \bibinfo{person}{Aiswarya Raj}, \bibinfo{person}{Jan Bosch},
  \bibinfo{person}{Helena~Holmstr{\"o}m Olsson}, {and} \bibinfo{person}{Ivica
  Crnkovic}.} \bibinfo{year}{2019}\natexlab{}.
\newblock \showarticletitle{A taxonomy of software engineering challenges for
  machine learning systems: An empirical investigation}. In
  \bibinfo{booktitle}{\emph{International Conference on Agile Software
  Development}}. Springer, \bibinfo{pages}{227--243}.
\newblock


\bibitem[\protect\citeauthoryear{Mathis, Avdiienko, Soremekun, B{\"o}hme, and
  Zeller}{Mathis et~al\mbox{.}}{2017}]%
        {mathis2017detecting}
\bibfield{author}{\bibinfo{person}{Bj{\"o}rn Mathis}, \bibinfo{person}{Vitalii
  Avdiienko}, \bibinfo{person}{Ezekiel~O Soremekun}, \bibinfo{person}{Marcel
  B{\"o}hme}, {and} \bibinfo{person}{Andreas Zeller}.}
  \bibinfo{year}{2017}\natexlab{}.
\newblock \showarticletitle{Detecting information flow by mutating input data}.
  In \bibinfo{booktitle}{\emph{2017 32nd IEEE/ACM International Conference on
  Automated Software Engineering (ASE)}}. IEEE, \bibinfo{pages}{263--273}.
\newblock


\bibitem[\protect\citeauthoryear{Menzies, Kocaguneli, Turhan, Minku, and
  Peters}{Menzies et~al\mbox{.}}{2014}]%
        {menzies2014sharing}
\bibfield{author}{\bibinfo{person}{Tim Menzies}, \bibinfo{person}{Ekrem
  Kocaguneli}, \bibinfo{person}{Burak Turhan}, \bibinfo{person}{Leandro Minku},
  {and} \bibinfo{person}{Fayola Peters}.} \bibinfo{year}{2014}\natexlab{}.
\newblock \bibinfo{booktitle}{\emph{Sharing data and models in software
  engineering}}.
\newblock \bibinfo{publisher}{Morgan Kaufmann}.
\newblock


\bibitem[\protect\citeauthoryear{Menzies and Zimmermann}{Menzies and
  Zimmermann}{2018}]%
        {menzies2018software}
\bibfield{author}{\bibinfo{person}{Tim Menzies} {and} \bibinfo{person}{Thomas
  Zimmermann}.} \bibinfo{year}{2018}\natexlab{}.
\newblock \showarticletitle{Software Analytics: What's Next?}
\newblock \bibinfo{journal}{\emph{IEEE Software}} \bibinfo{volume}{35},
  \bibinfo{number}{5} (\bibinfo{year}{2018}), \bibinfo{pages}{64--70}.
\newblock


\bibitem[\protect\citeauthoryear{Miller}{Miller}{2005}]%
        {MILLER2005233}
\bibfield{author}{\bibinfo{person}{James Miller}.}
  \bibinfo{year}{2005}\natexlab{}.
\newblock \showarticletitle{Replicating software engineering experiments: a
  poisoned chalice or the Holy Grail}.
\newblock \bibinfo{journal}{\emph{Information and Software Technology}}
  \bibinfo{volume}{47}, \bibinfo{number}{4} (\bibinfo{year}{2005}),
  \bibinfo{pages}{233 -- 244}.
\newblock
\showISSN{0950-5849}
\urldef\tempurl%
\url{https://doi.org/10.1016/j.infsof.2004.08.005}
\showDOI{\tempurl}


\bibitem[\protect\citeauthoryear{Misirli, Erdogmus, Juristo, and
  Dieste}{Misirli et~al\mbox{.}}{2014}]%
        {misirli2014topic}
\bibfield{author}{\bibinfo{person}{Ayse~Tosun Misirli}, \bibinfo{person}{Hakan
  Erdogmus}, \bibinfo{person}{Natalia Juristo}, {and} \bibinfo{person}{Oscar
  Dieste}.} \bibinfo{year}{2014}\natexlab{}.
\newblock \showarticletitle{Topic selection in industry experiments}. In
  \bibinfo{booktitle}{\emph{Proceedings of the 2nd International Workshop on
  Conducting Empirical Studies in Industry}}. \bibinfo{pages}{25--30}.
\newblock


\bibitem[\protect\citeauthoryear{Nayebi, Cai, Kazman, Ruhe, Feng, Carlson, and
  Chew}{Nayebi et~al\mbox{.}}{2019}]%
        {nayebi2019longitudinal}
\bibfield{author}{\bibinfo{person}{Maleknaz Nayebi}, \bibinfo{person}{Yuanfang
  Cai}, \bibinfo{person}{Rick Kazman}, \bibinfo{person}{Guenther Ruhe},
  \bibinfo{person}{Qiong Feng}, \bibinfo{person}{Chris Carlson}, {and}
  \bibinfo{person}{Francis Chew}.} \bibinfo{year}{2019}\natexlab{}.
\newblock \showarticletitle{A longitudinal study of identifying and paying down
  architecture debt}. In \bibinfo{booktitle}{\emph{2019 IEEE/ACM 41st
  International Conference on Software Engineering: Software Engineering in
  Practice (ICSE-SEIP)}}. IEEE, \bibinfo{pages}{171--180}.
\newblock


\bibitem[\protect\citeauthoryear{Nayebi, Kuznetsov, Chen, Zeller, and
  Ruhe}{Nayebi et~al\mbox{.}}{2018}]%
        {nayebi2018anatomy}
\bibfield{author}{\bibinfo{person}{Maleknaz Nayebi},
  \bibinfo{person}{Konstantin Kuznetsov}, \bibinfo{person}{Paul Chen},
  \bibinfo{person}{Andreas Zeller}, {and} \bibinfo{person}{Guenther Ruhe}.}
  \bibinfo{year}{2018}\natexlab{}.
\newblock \showarticletitle{Anatomy of functionality deletion: an exploratory
  study on mobile apps}. In \bibinfo{booktitle}{\emph{Proceedings of the 15th
  International Conference on Mining Software Repositories}}.
  \bibinfo{pages}{243--253}.
\newblock


\bibitem[\protect\citeauthoryear{Nayebi, Marbouti, Quapp, Maurer, and
  Ruhe}{Nayebi et~al\mbox{.}}{2017}]%
        {nayebi2017crowdsourced}
\bibfield{author}{\bibinfo{person}{Maleknaz Nayebi}, \bibinfo{person}{Mahshid
  Marbouti}, \bibinfo{person}{Rache Quapp}, \bibinfo{person}{Frank Maurer},
  {and} \bibinfo{person}{Guenther Ruhe}.} \bibinfo{year}{2017}\natexlab{}.
\newblock \showarticletitle{Crowdsourced exploration of mobile app features: A
  case study of the fort mcmurray wildfire}. In \bibinfo{booktitle}{\emph{2017
  IEEE/ACM 39th International Conference on Software Engineering: Software
  Engineering in Society Track (ICSE-SEIS)}}. IEEE, \bibinfo{pages}{57--66}.
\newblock


\bibitem[\protect\citeauthoryear{Roehm}{Roehm}{2015}]%
        {roehm2015two}
\bibfield{author}{\bibinfo{person}{Tobias Roehm}.}
  \bibinfo{year}{2015}\natexlab{}.
\newblock \showarticletitle{Two user perspectives in program comprehension: end
  users and developer users}. In \bibinfo{booktitle}{\emph{2015 IEEE 23rd
  International Conference on Program Comprehension}}. IEEE,
  \bibinfo{pages}{129--139}.
\newblock


\bibitem[\protect\citeauthoryear{Sarkar and Parnin}{Sarkar and Parnin}{2017}]%
        {sarkar2017characterizing}
\bibfield{author}{\bibinfo{person}{Saurabh Sarkar} {and} \bibinfo{person}{Chris
  Parnin}.} \bibinfo{year}{2017}\natexlab{}.
\newblock \showarticletitle{Characterizing and predicting mental fatigue during
  programming tasks}. In \bibinfo{booktitle}{\emph{2017 IEEE/ACM 2nd
  International Workshop on Emotion Awareness in Software Engineering
  (SEmotion)}}. IEEE, \bibinfo{pages}{32--37}.
\newblock


\bibitem[\protect\citeauthoryear{Sawant, Robbes, and Bacchelli}{Sawant
  et~al\mbox{.}}{2016}]%
        {sawant2016reaction}
\bibfield{author}{\bibinfo{person}{Anand~Ashok Sawant}, \bibinfo{person}{Romain
  Robbes}, {and} \bibinfo{person}{Alberto Bacchelli}.}
  \bibinfo{year}{2016}\natexlab{}.
\newblock \showarticletitle{On the reaction to deprecation of 25,357 clients of
  4+ 1 popular Java APIs}. In \bibinfo{booktitle}{\emph{2016 IEEE International
  Conference on Software Maintenance and Evolution (ICSME)}}. IEEE,
  \bibinfo{pages}{400--410}.
\newblock


\bibitem[\protect\citeauthoryear{Sawant, Robbes, and Bacchelli}{Sawant
  et~al\mbox{.}}{2018}]%
        {sawant2018reaction}
\bibfield{author}{\bibinfo{person}{Anand~Ashok Sawant}, \bibinfo{person}{Romain
  Robbes}, {and} \bibinfo{person}{Alberto Bacchelli}.}
  \bibinfo{year}{2018}\natexlab{}.
\newblock \showarticletitle{On the reaction to deprecation of clients of 4+ 1
  popular Java APIs and the JDK}.
\newblock \bibinfo{journal}{\emph{Empirical Software Engineering}}
  \bibinfo{volume}{23}, \bibinfo{number}{4} (\bibinfo{year}{2018}),
  \bibinfo{pages}{2158--2197}.
\newblock


\bibitem[\protect\citeauthoryear{Schermann, Sch{\"o}ni, Leitner, and
  Gall}{Schermann et~al\mbox{.}}{2016}]%
        {schermann2016bifrost}
\bibfield{author}{\bibinfo{person}{Gerald Schermann}, \bibinfo{person}{Dominik
  Sch{\"o}ni}, \bibinfo{person}{Philipp Leitner}, {and}
  \bibinfo{person}{Harald~C Gall}.} \bibinfo{year}{2016}\natexlab{}.
\newblock \showarticletitle{Bifrost: Supporting continuous deployment with
  automated enactment of multi-phase live testing strategies}. In
  \bibinfo{booktitle}{\emph{Proceedings of the 17th International Middleware
  Conference}}. \bibinfo{pages}{1--14}.
\newblock


\bibitem[\protect\citeauthoryear{Sharma, Mehra, and Kaulgud}{Sharma
  et~al\mbox{.}}{2017}]%
        {sharma2017developers}
\bibfield{author}{\bibinfo{person}{Vibhu~Saujanya Sharma},
  \bibinfo{person}{Rohit Mehra}, {and} \bibinfo{person}{Vikrant Kaulgud}.}
  \bibinfo{year}{2017}\natexlab{}.
\newblock \showarticletitle{What do developers want? an advisor approach for
  developer priorities}. In \bibinfo{booktitle}{\emph{2017 IEEE/ACM 10th
  International Workshop on Cooperative and Human Aspects of Software
  Engineering (CHASE)}}. IEEE, \bibinfo{pages}{78--81}.
\newblock


\bibitem[\protect\citeauthoryear{Shull, Carver, Vegas, and Juristo}{Shull
  et~al\mbox{.}}{2008}]%
        {Shull2008}
\bibfield{author}{\bibinfo{person}{Forrest~J. Shull},
  \bibinfo{person}{Jeffrey~C. Carver}, \bibinfo{person}{Sira Vegas}, {and}
  \bibinfo{person}{Natalia Juristo}.} \bibinfo{year}{2008}\natexlab{}.
\newblock \showarticletitle{The role of replications in Empirical Software
  Engineering}.
\newblock \bibinfo{journal}{\emph{Empirical Software Engineering}}
  \bibinfo{volume}{13}, \bibinfo{number}{2} (\bibinfo{date}{01 Apr}
  \bibinfo{year}{2008}), \bibinfo{pages}{211--218}.
\newblock
\showISSN{1573-7616}
\urldef\tempurl%
\url{https://doi.org/10.1007/s10664-008-9060-1}
\showDOI{\tempurl}


\bibitem[\protect\citeauthoryear{Storey, Zagalsky, Figueira~Filho, Singer, and
  German}{Storey et~al\mbox{.}}{2016}]%
        {storey2016social}
\bibfield{author}{\bibinfo{person}{Margaret-Anne Storey},
  \bibinfo{person}{Alexey Zagalsky}, \bibinfo{person}{Fernando Figueira~Filho},
  \bibinfo{person}{Leif Singer}, {and} \bibinfo{person}{Daniel~M German}.}
  \bibinfo{year}{2016}\natexlab{}.
\newblock \showarticletitle{How social and communication channels shape and
  challenge a participatory culture in software development}.
\newblock \bibinfo{journal}{\emph{IEEE Transactions on Software Engineering}}
  \bibinfo{volume}{43}, \bibinfo{number}{2} (\bibinfo{year}{2016}),
  \bibinfo{pages}{185--204}.
\newblock


\bibitem[\protect\citeauthoryear{Suonsyrj{\"a}, Hokkanen, Terho, Syst{\"a}, and
  Mikkonen}{Suonsyrj{\"a} et~al\mbox{.}}{2016}]%
        {suonsyrja2016post}
\bibfield{author}{\bibinfo{person}{Sampo Suonsyrj{\"a}}, \bibinfo{person}{Laura
  Hokkanen}, \bibinfo{person}{Henri Terho}, \bibinfo{person}{Kari Syst{\"a}},
  {and} \bibinfo{person}{Tommi Mikkonen}.} \bibinfo{year}{2016}\natexlab{}.
\newblock \showarticletitle{Post-deployment data: A recipe for satisfying
  knowledge needs in software development?}. In \bibinfo{booktitle}{\emph{2016
  Joint Conference of the International Workshop on Software Measurement and
  the International Conference on Software Process and Product Measurement
  (IWSM-MENSURA)}}. IEEE, \bibinfo{pages}{139--147}.
\newblock


\bibitem[\protect\citeauthoryear{Suonsyrj{\"a} and Mikkonen}{Suonsyrj{\"a} and
  Mikkonen}{2015}]%
        {suonsyrja2015designing}
\bibfield{author}{\bibinfo{person}{Sampo Suonsyrj{\"a}} {and}
  \bibinfo{person}{Tommi Mikkonen}.} \bibinfo{year}{2015}\natexlab{}.
\newblock \showarticletitle{Designing an unobtrusive analytics framework for
  monitoring java applications}.
\newblock In \bibinfo{booktitle}{\emph{Software Measurement}}.
  \bibinfo{publisher}{Springer}, \bibinfo{pages}{160--175}.
\newblock


\bibitem[\protect\citeauthoryear{Tavakoli, Zhao, Heydari, and
  Nenadi{\'c}}{Tavakoli et~al\mbox{.}}{2018}]%
        {tavakoli2018extracting}
\bibfield{author}{\bibinfo{person}{Mohammadali Tavakoli},
  \bibinfo{person}{Liping Zhao}, \bibinfo{person}{Atefeh Heydari}, {and}
  \bibinfo{person}{Goran Nenadi{\'c}}.} \bibinfo{year}{2018}\natexlab{}.
\newblock \showarticletitle{Extracting useful software development information
  from mobile application reviews: A survey of intelligent mining techniques
  and tools}.
\newblock \bibinfo{journal}{\emph{Expert Systems with Applications}}
  \bibinfo{volume}{113} (\bibinfo{year}{2018}), \bibinfo{pages}{186--199}.
\newblock


\bibitem[\protect\citeauthoryear{Tripathi, Dabral, and Sureka}{Tripathi
  et~al\mbox{.}}{2015}]%
        {tripathi2015university}
\bibfield{author}{\bibinfo{person}{Ambika Tripathi}, \bibinfo{person}{Savita
  Dabral}, {and} \bibinfo{person}{Ashish Sureka}.}
  \bibinfo{year}{2015}\natexlab{}.
\newblock \showarticletitle{University-industry collaboration and open source
  software (oss) dataset in mining software repositories (msr) research}. In
  \bibinfo{booktitle}{\emph{2015 IEEE 1st International Workshop on Software
  Analytics (SWAN)}}. IEEE, \bibinfo{pages}{39--40}.
\newblock


\bibitem[\protect\citeauthoryear{Vassallo, Zampetti, Romano, Beller,
  Panichella, Di~Penta, and Zaidman}{Vassallo et~al\mbox{.}}{2016}]%
        {vassallo2016continuous}
\bibfield{author}{\bibinfo{person}{Carmine Vassallo}, \bibinfo{person}{Fiorella
  Zampetti}, \bibinfo{person}{Daniele Romano}, \bibinfo{person}{Moritz Beller},
  \bibinfo{person}{Annibale Panichella}, \bibinfo{person}{Massimiliano
  Di~Penta}, {and} \bibinfo{person}{Andy Zaidman}.}
  \bibinfo{year}{2016}\natexlab{}.
\newblock \showarticletitle{Continuous delivery practices in a large financial
  organization}. In \bibinfo{booktitle}{\emph{2016 IEEE International
  Conference on Software Maintenance and Evolution (ICSME)}}. IEEE,
  \bibinfo{pages}{519--528}.
\newblock


\bibitem[\protect\citeauthoryear{Wan, Xia, Hassan, Lo, Yin, and Yang}{Wan
  et~al\mbox{.}}{2018}]%
        {wan2018perceptions}
\bibfield{author}{\bibinfo{person}{Zhiyuan Wan}, \bibinfo{person}{Xin Xia},
  \bibinfo{person}{Ahmed~E Hassan}, \bibinfo{person}{David Lo},
  \bibinfo{person}{Jianwei Yin}, {and} \bibinfo{person}{Xiaohu Yang}.}
  \bibinfo{year}{2018}\natexlab{}.
\newblock \showarticletitle{Perceptions, expectations, and challenges in defect
  prediction}.
\newblock \bibinfo{journal}{\emph{IEEE Transactions on Software Engineering}}
  (\bibinfo{year}{2018}).
\newblock


\bibitem[\protect\citeauthoryear{Widder, Hilton, K{\"a}stner, and
  Vasilescu}{Widder et~al\mbox{.}}{2019}]%
        {widder2019conceptual}
\bibfield{author}{\bibinfo{person}{David~Gray Widder}, \bibinfo{person}{Michael
  Hilton}, \bibinfo{person}{Christian K{\"a}stner}, {and}
  \bibinfo{person}{Bogdan Vasilescu}.} \bibinfo{year}{2019}\natexlab{}.
\newblock \showarticletitle{A conceptual replication of continuous integration
  pain points in the context of Travis CI}. In
  \bibinfo{booktitle}{\emph{Proceedings of the 2019 27th ACM Joint Meeting on
  European Software Engineering Conference and Symposium on the Foundations of
  Software Engineering}}. \bibinfo{pages}{647--658}.
\newblock


\bibitem[\protect\citeauthoryear{Wohlin}{Wohlin}{2014}]%
        {Wohlin:2014:GSS:2601248.2601268}
\bibfield{author}{\bibinfo{person}{Claes Wohlin}.}
  \bibinfo{year}{2014}\natexlab{}.
\newblock \showarticletitle{Guidelines for Snowballing in Systematic Literature
  Studies and a Replication in Software Engineering}. In
  \bibinfo{booktitle}{\emph{Proceedings of the 18th International Conference on
  Evaluation and Assessment in Software Engineering}}
  \emph{(\bibinfo{series}{EASE '14})}. \bibinfo{publisher}{ACM},
  \bibinfo{address}{New York, NY, USA}, Article \bibinfo{articleno}{38},
  \bibinfo{numpages}{10}~pages.
\newblock
\showISBNx{978-1-4503-2476-2}
\urldef\tempurl%
\url{https://doi.org/10.1145/2601248.2601268}
\showDOI{\tempurl}


\bibitem[\protect\citeauthoryear{Zampetti, Scalabrino, Oliveto, Canfora, and
  Di~Penta}{Zampetti et~al\mbox{.}}{2017}]%
        {zampetti2017open}
\bibfield{author}{\bibinfo{person}{Fiorella Zampetti}, \bibinfo{person}{Simone
  Scalabrino}, \bibinfo{person}{Rocco Oliveto}, \bibinfo{person}{Gerardo
  Canfora}, {and} \bibinfo{person}{Massimiliano Di~Penta}.}
  \bibinfo{year}{2017}\natexlab{}.
\newblock \showarticletitle{How open source projects use static code analysis
  tools in continuous integration pipelines}. In \bibinfo{booktitle}{\emph{2017
  IEEE/ACM 14th International Conference on Mining Software Repositories
  (MSR)}}. IEEE, \bibinfo{pages}{334--344}.
\newblock


\bibitem[\protect\citeauthoryear{Zimmermann}{Zimmermann}{2017}]%
        {Zimmermann:2017:SPD:3084100.3087674}
\bibfield{author}{\bibinfo{person}{Thomas Zimmermann}.}
  \bibinfo{year}{2017}\natexlab{}.
\newblock \showarticletitle{Software Productivity Decoded: How Data Science
  Helps to Achieve More (Keynote)}. In \bibinfo{booktitle}{\emph{Proceedings of
  the 2017 International Conference on Software and System Process}}
  \emph{(\bibinfo{series}{ICSSP 2017})}. \bibinfo{publisher}{ACM},
  \bibinfo{address}{New York, NY, USA}, \bibinfo{pages}{1--2}.
\newblock
\showISBNx{978-1-4503-5270-3}
\urldef\tempurl%
\url{https://doi.org/10.1145/3084100.3087674}
\showDOI{\tempurl}


\bibitem[\protect\citeauthoryear{Zou, Lo, Chen, Xia, Feng, and Xu}{Zou
  et~al\mbox{.}}{2018}]%
        {zou2018practitioners}
\bibfield{author}{\bibinfo{person}{Weiqin Zou}, \bibinfo{person}{David Lo},
  \bibinfo{person}{Zhenyu Chen}, \bibinfo{person}{Xin Xia},
  \bibinfo{person}{Yang Feng}, {and} \bibinfo{person}{Baowen Xu}.}
  \bibinfo{year}{2018}\natexlab{}.
\newblock \showarticletitle{How practitioners perceive automated bug report
  management techniques}.
\newblock \bibinfo{journal}{\emph{IEEE Transactions on Software Engineering}}
  (\bibinfo{year}{2018}).
\newblock


\end{thebibliography}

\onecolumn
\clearpage
\section*{Appendix}
\label{techreport}
This appendix contains methodological and statistical supplements for the paper 'Questions for Data Scientists in Software Engineering'. The report gives in Table \ref{tab:category_mapping} and overview of the mapping of categories from the \ING 2019 study on the categories in the original Microsoft 2014 study. Table \ref{tab:all_descriptive_questions} includes an overview of all 171 descriptive questions that resulted from the initial survey, and that were used for ranking purposes in the ranking survey. The table is sorted on Percentages Worthwhile+ and Percentages Essential.
Furthermore, Table \ref{tab:demographics_discipline_all} gives an overview of statistically significant rating differences by demographics.

\begin{table} [ht]
  \caption{Category mapping.}
  \label{tab:category_mapping}
  \footnotesize
    \begin{tabularx}{\linewidth}{
    >{\hsize=0.7\hsize}X
    >{\hsize=0.25\hsize}X
    >{\hsize=0.7\hsize}X
    >{\hsize=0.25\hsize}X
    }
    \toprule
    Microsoft Category & Descriptive & \ING Category & Descriptive\\
     & Questions & & Questions\\
    \midrule
Development & 28 & Estimation & 8 \\
Practices &  & Architecture & 7 \\
 &  & Knowledge sharing & 3 \\
 &  & Dependencies & 3 \\
 \midrule
Testing Practices & 20 & Testing & 13 \\
 \midrule
Evaluating Quality & 16 & Code analysis & 5 \\
 &  & Quality & 5 \\
 &  & Effectiveness & 1 \\
 \midrule
Software & 14 & Development processes & 17 \\
Development &  & Technology selection & 3 \\
Process &  & Deployment & 2 \\
 \midrule
Productivity & 13 & Productivity & 8 \\
 &  & Employee evaluation & 6 \\
 \midrule
Teams and & 11 & Team composition & 3 \\
Collaboration &  & Knowledge sharing & 2 \\
 \midrule
Customers and & 9 & Formal methods & 4 \\
Requirements &  & Customer value & 3 \\
 \midrule
Development & 9 & Best practices & 11 \\
Best Practices &  & Technology selection & 7 \\
 &  & Effectiveness & 7 \\
 &  & Tools & 5 \\
 \midrule
Services & 8 & Performance & 1 \\
 \midrule
Bug Measurements & 7 & Security & 3 \\
 \midrule
Software & 7 & Effectiveness & 1 \\
Development &  & Management & 1 \\
Lifecycle &  & Performance & 1 \\
 &  & Security & 1 \\
 \midrule
Reuse and & 3 & Awareness & 1 \\
Shared Components &  & Reuse & 1 \\
    \bottomrule
    \\[-4pt]
    \end{tabularx}
    \begin{flushleft}
    \end{flushleft}
\end{table}

\begin{table*}[h]
\caption{Overview of all descriptive questions.}
\label{tab:all_descriptive_questions}
\small
\begin{tabular}{p{0.4cm}p{7.8cm}p{1.0cm}|P{0.8cm}P{1.2cm}p{0.7cm}|P{0.8cm}P{1.2cm}P{0.7cm}}
\toprule
 &  &  & \multicolumn{3}{c}{Percentages} & \multicolumn{3}{c}{Rank} \\
 & Question & Category & Essential & Worthwhile+ & Unwise & Essential & Worthwhile+ & Unwise \\ \midrule
Q143 & What factors affect the performance and productivity of DevOps teams with regard to team happiness and pleasure in your work? & DP & 68.4\% & 94.7\% & 0.0\% & 1 & 9 & 68 \\
Q98 & To what extent does on-demand provisioning of develop- and test environments, including up-to-date data affect delivery of software solutions? & TP & 66.7\% & 77.8\% & 0.0\% & 2 & 95 & 68 \\
Q37 & How can we make sure that we build for re-usability and scalability? & BEST & 63.2\% & 89.5\% & 5.3\% & 3 & 42 & 63 \\
Q145 & What factors affect the performance of DevOps teams and the quality of software code with regard to quantity and quality of environments? & PROD & 60.0\% & 100.0\% & 0.0\% & 4 & 1 & 68 \\
Q114 & What factors affect High Performance Teams? & BEST & 58.8\% & 82.4\% & 0.0\% & 5 & 75 & 68 \\
\midrule
Q154 & What factors affect understand-ability and readability of software code for other developers? & DP & 58.3\% & 91.7\% & 8.3\% & 6 & 25 & 44 \\
Q76 & To what extent affects building software solutions by using a continuous delivery pipeline with automated testing and migrating, including rollback facilities the performance of DevOps teams and the quality of software? & PROC & 56.3\% & 93.8\% & 0.0\% & 7 & 15 & 68 \\
Q36 & How can we improve the deployment process in DevOps teams? & PROC & 56.3\% & 93.8\% & 0.0\% & 7 & 15 & 68 \\
Q53 & How does the effort spent on fixing vulnerabilities and bugs relate to effort spent on writing software correctly from the start? & BUG & 56.3\% & 93.8\% & 0.0\% & 7 & 15 & 68 \\
Q22 & How can requirements be validated before starting actual software development? & CR & 55.6\% & 88.9\% & 0.0\% & 10 & 44 & 68 \\
\midrule
Q123 & What factors affect performance testing on high data volumes? & TP & 55.6\% & 88.9\% & 0.0\% & 10 & 44 & 68 \\
Q58 & How to measure the customer value of a software product? & CR & 55.6\% & 77.8\% & 11,1\% & 10 & 95 & 20 \\
Q163 & What methods can be applied to analyze whether software code is working as expected? & EQ & 53.3\% & 93.3\% & 0.0\% & 13 & 19 & 68 \\
Q88 & To what extent affects the test ability of software code the quality of code? & EQ & 52.9\% & 100.0\% & 0.0\% & 14 & 1 & 68 \\
Q80 & To what extent affects implementing automated controls within the continuous delivery pipeline the effort spent on accountability with regard to risks and security? & DP & 50.0\% & 95.0\% & 0.0\% & 15 & 8 & 68 \\
\midrule
Q125 & What factors affect providing new technologies to consumers, and can implementations of new technology be internally and externally benchmarked? & SL & 50.0\% & 90.0\% & 0.0\% & 15 & 30 & 68 \\
Q109 & What factors affect creating and maintaining software solutions yourself, versus using off-the-shelve solutions? & BEST & 50.0\% & 90.0\% & 0.0\% & 15 & 30 & 68 \\
Q34 & How can user feedback be integrated in an efficient and effective way into software code? & CR & 50.0\% & 87.5\% & 12.5\% & 15 & 56 & 15 \\
Q1 & Are developers working in an open space with several teams more effective or less than developers working in a room with just their team? & PROD & 50.0\% & 83.3\% & 8.3\% & 15 & 73 & 44 \\
Q140 & What factors affect the level of competence - such as novice, advanced, competent - of software engineers in order to gain insight into tasks, time spent, and development goals per competence level. & DP & 50.0\% & 75.0\% & 8.3\% & 15 & 106 & 44 \\
\bottomrule
\end{tabular}
\end{table*}

\begin{table*}[h]
\label{tab:all_descriptive_questions2}
\small
\begin{tabular}{p{0.4cm}p{7.8cm}p{1.0cm}|P{0.8cm}P{1.2cm}p{0.7cm}|P{0.8cm}P{1.2cm}P{0.7cm}}
\toprule
 &  &  & \multicolumn{3}{c}{Percentages} & \multicolumn{3}{c}{Rank} \\
 & Question & Category & Essential & Worthwhile+ & Unwise & Essential & Worthwhile+ & Unwise \\ \midrule
Q159 & What is the relation between frameworks used and programming languages one hand and time of development on  the other? & DP & 50.0\% & 75.0\% & 0.0\% & 15 & 106 & 68 \\
Q8 & Does a focus on time and effort affects the quality of software systems in terms of availability, reliability, and security? & RSC & 50.0\% & 70.0\% & 0.0\% & 15 & 122 & 68 \\
Q33 & How can toolsets be organized into distributable packages, in order to support installation of components by software developers? & BEST & 47.4\% & 84.2\% & 0.0\% & 23 & 70 & 68 \\
Q166 & When do you remove an old module that you think is not being used anymore? & BEST & 47.4\% & 73.7\% & 0.0\% & 23 & 113 & 68 \\
Q67 & To what extent affects automated checks of coding conventions, code quality, code complexity, and  test-coverage the quality of software systems and the performance of DevOps teams? & EQ & 47.1\% & 100.0\% & 0.0\% & 25 & 1 & 68 \\
\midrule
Q82 & To what extent affects peer review or code review the quality of code? & EQ & 47.1\% & 94.1\% & 0.0\% & 25 & 12 & 68 \\
Q75 & To what extent affects automation of deployment scripts, with regard to single use versus re-use, the delivery of software solutions? & PROC & 46.7\% & 86.7\% & 0.0\% & 27 & 59 & 68 \\
Q162 & What methods are most effective in preventing security related vulnerabilities or bugs from being introduced  in software code? & BUG & 46.7\% & 86.7\% & 0.0\% & 27 & 59 & 68 \\
Q7 & Does a focus on quick release of  features and bug fixes into production help to achieve confidence and   agility? & PROC & 46.7\% & 80.0\% & 6.7\% & 27 & 84 & 51 \\
Q141 & What factors affect the   performance and productivity of DevOps teams with regard to evidence-based   decision-making versus decision-making based on expert opinions. & DP & 45.5\% & 90.9\% & 0.0\% & 30 & 26 & 68 \\
\midrule
Q32 & How can the productivity of   engineers and squads be measured? & PROD & 45.5\% & 54.5\% & 9.1\% & 30 & 155 & 36 \\
Q11 & How can a system for (semi)   automated CRUD test data generation improve delivery of software solutions? & TP & 44.4\% & 100.0\% & 0.0\% & 32 & 1 & 68 \\
Q41 & How do customers perceive   software interfaces with regard to clarity and fitness of a solution? & CR & 44.4\% & 88.9\% & 11.1\% & 32 & 44 & 20 \\
Q73 & To what extent affects a test   coverage approach, including a specific threshold, the delivery and quality   of software solutions? & TP & 44.4\% & 77.8\% & 0.0\% & 32 & 95 & 68 \\
Q118 & What factors affect integration   testing of software solutions? & TP & 44.4\% & 77.8\% & 0.0\% & 32 & 95 & 68 \\
\midrule
Q161 & What makes a great coder? What   aspects affect the performance of DevOps teams and the quality of software   with regard to characteristics of an individual software engineer? & PROD & 44.4\% & 66.7\% & 22.2\% & 32 & 128 & 4 \\
Q111 & What factors affect debugging of   complex data issues in production? & TP & 44.4\% & 66.7\% & 0.0\% & 32 & 128 & 68 \\
Q64 & Is the infrastructure for   development and in the OTAP flexible and modern enough to ease and speed up   development? & PROC & 43.8\% & 81.3\% & 12.5\% & 38 & 82 & 15 \\
Q92 & To what extent can an approach   such as Chaos Monkey, where virtual machine instances and containers that run   in the production environment are randomly terminated, help to build more   resilient software services? & TP & 42.9\% & 85.7\% & 0.0\% & 39 & 64 & 68 \\
Q14 & How can an inventory of tips and  tricks for software development help software engineers to develop software? & BEST & 42.1\% & 94.7\% & 0.0\% & 40 & 9 & 68 \\
\midrule
Q120 & What factors affect limitation   of access by engineers to production systems versus effectiveness of   engineers? & BEST & 42.1\% & 78.9\% & 10.5\% & 40 & 93 & 29 \\
Q95 & To what extent do dependencies   on other teams affect team performance? & TC & 41.2\% & 94.1\% & 0.0\% & 42 & 12 & 68 \\
Q121 & What factors affect loading high   volume data with regard to availability of tools in industry? & BEST & 41.2\% & 76.5\% & 0.0\% & 42 & 105 & 68 \\
Q104 & What aspects affects the   performance of DevOps teams and the quality of software with regard to   software architecture? & PROD & 40.0\% & 100.0\% & 0.0\% & 44 & 1 & 68 \\
Q164 & What parts of a software system   affect software quality most, with regard to aspects such as change   frequency, bug severity, hit ratio of paths, and importance of code? & EQ & 40.0\% & 93.3\% & 0.0\% & 44 & 19 & 68 \\
\bottomrule
\end{tabular}
\end{table*}

\begin{table*}[h]
\label{tab:all_descriptive_questions3}
\small
\begin{tabular}{p{0.4cm}p{7.8cm}p{1.0cm}|P{0.8cm}P{1.2cm}p{0.7cm}|P{0.8cm}P{1.2cm}P{0.7cm}}
\toprule
 &  &  & \multicolumn{3}{c}{Percentages} & \multicolumn{3}{c}{Rank} \\
 & Question & Category & Essential & Worthwhile+ & Unwise & Essential & Worthwhile+ & Unwise \\ \midrule
Q47 & How do the number and severity   of security incidents relate to platforms or types of software such as Linux,   Windows, Mainframe, CDaaS, non-CDaaS, TFS, BOTS, SAAS, and inhouse   development? & BUG & 40.0\% & 93.3\% & 0.0\% & 44 & 19 & 68 \\
Q48 & How does automated testing as   part of the design and build process influence the time of delivery of   software products? & BEST & 40.0\% & 90.0\% & 0.0\% & 44 & 30 & 68 \\
Q70 & To what extent affect the   requirements of a laptop the performance of DevOps teams and the quality of   software? & PROD & 40.0\% & 90.0\% & 0.0\% & 44 & 30 & 68 \\
Q113 & What factors affect estimation   of lead time, cost, and quality of software deliveries? & SL & 40.0\% & 80.0\% & 0.0\% & 44 & 84 & 68 \\
Q85 & To what extent affects security   by design - meaning that the software has been designed from the foundation   to be secure - the delivery of software solutions? & SL & 40.0\% & 80.0\% & 0.0\% & 44 & 84 & 68 \\
\midrule
Q133 & What factors affect testing and   fixing bugs during a sprint versus testing and fixing bugs just before   release (after several sprints)? & BEST & 40.0\% & 80.0\% & 0.0\% & 44 & 84 & 68 \\
Q168 & Which input data is in the end   result worth say ten times more than other? & TP & 40.0\% & 80.0\% & 0.0\% & 44 & 84 & 68 \\
Q30 & How can the performance of   individual software engineers be benchmarked internally \ING and externally   with other companies? & PROD & 40.0\% & 50.0\% & 20.0\% & 44 & 157 & 6 \\
Q131 & What factors affect team   on-boarding on \ING frameworks with regard to speed of development? & DP & 38.9\% & 88.9\% & 0.0\% & 54 & 44 & 68 \\
Q12 & How can an inventory of best   practices for analytical solutions in software development help software   engineers to develop software? & BEST & 38.9\% & 88.9\% & 0.0\% & 54 & 44 & 68 \\
\midrule
Q165 & What properties affect the   quality of tooling for backlog management, development, such as tools for   Configuration Items, issue tracking, logging, monitoring, and crash   reporting? & BEST & 38.9\% & 88.9\% & 0.0\% & 54 & 44 & 68 \\
Q153 & What factors affect trunk-based   development - a source-control branching model, where developers collaborate   on code in a single branch - with regard to quality of software code? & BEST & 38.9\% & 83.3\% & 5.6\% & 54 & 73 & 61 \\
Q134 & What factors affect TFS (Team   Foundation Services) - a Microsoft product that provides source code   management - with regard to working with automated pipelines? & BEST & 38.9\% & 72.2\% & 22.2\% & 54 & 118 & 4 \\
Q63 & Is it more expensive to develop   front-end applications using 'modern' JavaScript frameworks, like Angular,   Polymer, React than frameworks like Vaadin, GWT, Icefaces and maybe some   others around? & BEST & 38.5\% & 84.6\% & 7.7\% & 59 & 69 & 47 \\
Q71 & To what extent affects a   checklist before pushing code into production improve performance and quality   of a software system? & DP & 38.1\% & 85.7\% & 0.0\% & 60 & 64 & 68 \\
\midrule
Q52 & How does team maturity affect   code quality and incidents? & TC & 37.5\% & 93.8\% & 0.0\% & 61 & 15 & 68 \\
Q142 & What factors affect the performance and productivity of DevOps teams with regard to simultaneous slow and fast developments at the same time in same environments? & DP & 37.5\% & 87.5\% & 0.0\% & 61 & 56 & 68 \\
Q139 & What factors affect the different software engineering approaches such as Scrum and Kanban in practical settings? & PROC & 37.5\% & 68.8\% & 12.5\% & 61 & 124 & 15 \\
Q127 & What factors affect replication of production data into a test environment in order to be able to test   migrations and releases on an exact copy of the production environment? & DP & 36.4\% & 90.9\% & 9.1\% & 64 & 26 & 36 \\
Q115 & What factors affect individual   software engineer productivity with regard to facilitation by others? & PROD & 36.4\% & 90.9\% & 9.1\% & 64 & 26 & 36 \\
\midrule
Q116 & What factors affect individual   software engineer productivity with regard to knowledge and skills? & PROD & 36.4\% & 81.8\% & 9.1\% & 64 & 77 & 36 \\
Q45 & How do software engineers deal with pressure from management or stakeholders? & PROD & 36.4\% & 81.8\% & 0.0\% & 64 & 77 & 68 \\
Q147 & What factors affect the requirements for a version control system in order to test for fitness within the \ING environment? & DP & 36.4\% & 72.7\% & 0.0\% & 64 & 115 & 68 \\
Q59 & How valuable would it be to have architects that are more busy with helping other developer and not being   fully busy with own development? & PROD & 36.4\% & 72.7\% & 0.0\% & 64 & 115 & 68 \\
Q135 & What factors affect the composition of DevOps teams? & TC & 35.7\% & 92.9\% & 0.0\% & 70 & 24 & 68 \\
\bottomrule
\end{tabular}
\end{table*}

\begin{table*}[h]
\label{tab:all_descriptive_questions4}
\small
\begin{tabular}{p{0.4cm}p{7.8cm}p{1.0cm}|P{0.8cm}P{1.2cm}p{0.7cm}|P{0.8cm}P{1.2cm}P{0.7cm}}
\toprule
 &  &  & \multicolumn{3}{c}{Percentages} & \multicolumn{3}{c}{Rank} \\
 & Question & Category & Essential & Worthwhile+ & Unwise & Essential & Worthwhile+ & Unwise \\ \midrule
Q96 & To what extent do standardization for solution development affect team performance? & TC & 35.3\% & 94.1\% & 0.0\% & 71 & 12 & 68 \\
Q94 & To what extent do dependencies of people, people changing roles or people leaving a team affect team  performance? & TC & 35.3\% & 64.7\% & 5.9\% & 71 & 138 & 59 \\
Q19 & How can editors help software developers to document their public functions in a way that it is available   for other developers? & CR & 33.3\% & 100.0\% & 0.0\% & 73 & 1 & 68 \\
Q122 & What factors affect  maintainability of software systems? & EQ & 33.3\% & 100.0\% & 0.0\% & 73 & 1 & 68 \\
Q10 & How are deadlines be handled   within the scope of an agile way of working? & PROC & 33.3\% & 88.9\% & 0.0\% & 73 & 44 & 68 \\
\midrule
Q119 & What factors affect lead time of   software deliveries, with regard to idea creation, design, development, test,   deploy, and release? & TP & 33.3\% & 88.9\% & 11.1\% & 73 & 44 & 20 \\
Q137 & What factors affect the delivery   of software solutions - including aspects such as architecture,   modularization, and distributed components - with regard to collaboration of   teams located in different countries? & PROC & 33.3\% & 86.7\% & 6.7\% & 73 & 59 & 51 \\
Q90 & To what extent affects the   upfront preparation of a design the delivery of software solutions? & PROC & 33.3\% & 86.7\% & 0.0\% & 73 & 59 & 68 \\
Q2 & Are there practices of good   software teams from the perspective of releasing software solutions into   production? & PROC & 33.3\% & 80.0\% & 0.0\% & 73 & 84 & 68 \\
Q107 & What factors affect adopting a   micro-service architecture - an architectural style that structures an   application as a collection of loosely coupled services, which implement   business capabilities - for software engineering purposes? & DP & 33.3\% & 77.8\% & 5.6\% & 73 & 95 & 61 \\
\midrule
Q18 & How can descriptive statistics   such as averages, minimum, and maximum of data-sets in use be measured in   order to scale the processing of data? & DP & 33.3\% & 77.8\% & 0.0\% & 73 & 95 & 68 \\
Q146 & What factors affect the quality   of the CDaaS solution for continuous delivery of software, and how can   improvements be realized? & PROC & 33.3\% & 77.8\% & 0.0\% & 73 & 95 & 68 \\
Q50 & How does involvement of business   stakeholders affect the delivery of software solutions? & TP & 33.3\% & 77.8\% & 11.1\% & 73 & 95 & 20 \\
Q100 & To what extent does Test Driven   Development - meaning that first test are designed and after that the code is   prepared - affects the delivery and quality of software solutions? & TP & 33.3\% & 77.8\% & 11.1\% & 73 & 95 & 20 \\
Q132 & What factors affect team   performance and system quality with regard to the number of teams working   simultaneously on one application? & DP & 33.3\% & 75.0\% & 0.0\% & 73 & 106 & 68 \\
\midrule
Q29 & How can the performance of   DevOps teams be benchmarked over departments and with regard to the Dreyfus   model? & PROD & 33.3\% & 66.7\% & 0.0\% & 73 & 128 & 68 \\
Q43 & How do naming standards affect   the development of software? & CR & 33.3\% & 66.7\% & 11.1\% & 73 & 128 & 20 \\
Q97 & To what extent does   documentation during software maintenance affect delivery of software   solutions? & TP & 33.3\% & 66.7\% & 11.1\% & 73 & 128 & 20 \\
Q152 & What factors affect the way how   DevOps teams perform, with regard to product ownership, and business   responsibilities? & DP & 31.6\% & 94.7\% & 0.0\% & 89 & 9 & 68 \\
Q69 & To what extent affect pair   programming - a software development technique in which two programmers work   together at one workstation and switch roles from code writer to reviewer   frequently - the delivery of software solutions? & PROC & 31.3\% & 75.0\% & 6.3\% & 90 & 106 & 55 \\
\midrule
Q74 & To what extent affects an agile   way of working the platform or tools used for delivery of software solutions? & PROC & 31.3\% & 68.8\% & 12.5\% & 90 & 124 & 15 \\
Q56 & How much time is spent on risk   management, security coding, and resilient architecture as part of software   development? & SL & 30.0\% & 90.0\% & 0.0\% & 92 & 30 & 68 \\
Q6 & Do unit tests save more time in   debugging than they take to write, run, or keep updated? & BEST & 30.0\% & 90.0\% & 0.0\% & 92 & 30 & 68 \\
Q129 & What factors affect running   systems on many Linux platforms versus running systems on a centralized   mainframe? & BEST & 30.0\% & 90.0\% & 0.0\% & 92 & 30 & 68 \\
Q55 & How much time does it take for a   beginning software engineer to start having a real contribution to a DevOps   team, and how to optimize that? & PROD & 30.0\% & 90.0\% & 0.0\% & 92 & 30 & 68 \\
\bottomrule
\end{tabular}
\end{table*}

\begin{table*}[h]
\label{tab:all_descriptive_questions5}
\small
\begin{tabular}{p{0.4cm}p{7.8cm}p{1.0cm}|P{0.8cm}P{1.2cm}p{0.7cm}|P{0.8cm}P{1.2cm}P{0.7cm}}
\toprule
 &  &  & \multicolumn{3}{c}{Percentages} & \multicolumn{3}{c}{Rank} \\
 & Question & Category & Essential & Worthwhile+ & Unwise & Essential & Worthwhile+ & Unwise \\ \midrule
Q103 & What aspects affects the   performance of DevOps teams and the quality of software with regard to an   agile way of working and working in DevOps teams? & PROD & 30.0\% & 90.0\% & 0.0\% & 92 & 30 & 68 \\
Q20 & How can effort needed - in terms   of time, effort, and cost - to perform a software project reliable be   estimated? & DP & 30.0\% & 80.0\% & 10.0\% & 92 & 84 & 32 \\
Q112 & What factors affect different   test frameworks with regard to effort spend and result of the test performed? & BEST & 30.0\% & 60.0\% & 0.0\% & 92 & 144 & 68 \\
Q148 & What factors affect the use of   Docker - combined with Docker management solution, such as kubernetes - and a   delivery pipeline, such as the one supplied by open-stack with regard to team   performance? & BEST & 28.6\% & 85.7\% & 7.1\% & 99 & 64 & 49 \\
Q144 & What factors affect the   performance of DevOps teams and the quality of software code with regard to   growing software systems organically versus start building a software system   from scratch? & EQ & 28.6\% & 85.7\% & 0.0\% & 99 & 64 & 68 \\
\midrule
Q150 & What factors affect the use of   PowerShell versus Ansible for deployment of systems? & BEST & 27.8\% & 55.6\% & 0.0\% & 101 & 152 & 68 \\
Q126 & What factors affect reliability   and security of software libraries? & DP & 27.3\% & 90.9\% & 0.0\% & 102 & 26 & 68 \\
Q17 & How can data be cashed in order   to prevent from retrieving data multiple times? & DP & 27.3\% & 81.8\% & 9.1\% & 102 & 77 & 36 \\
Q87 & To what extent affects the   reservation of time and effort dedicated for re-factoring purposes the   performance of a DevOps team and the quality of software? & PROD & 27.3\% & 81.8\% & 0.0\% & 102 & 77 & 68 \\
Q84 & To what extent affects screen   size the performance and productivity of software engineers? & PROD & 27.3\% & 81.8\% & 0.0\% & 102 & 77 & 68 \\
\midrule
Q26 & How can software engineers know   what other engineers are using a software component when adjusting one? & RSC & 27.3\% & 72.7\% & 0.0\% & 102 & 115 & 68 \\
Q83 & To what extent affects redesign   of code the quality of code and the performance of DevOps teams? & EQ & 26.7\% & 93.3\% & 0.0\% & 107 & 19 & 68 \\
Q136 & What factors affect the current   setup of networks and WIFI versus an open segment of the network where   Internet and a limited number of \ING network resources are available? & BEST & 26.7\% & 73.3\% & 0.0\% & 107 & 114 & 68 \\
Q31 & How can the process - in terms   of the state of individual requests - be monitored in the production   environment? & SVC & 26.3\% & 89.5\% & 0.0\% & 109 & 42 & 68 \\
Q108 & What factors affect an   architectural design of an application with regard to availability of such a   design, and quality of such a design? & DP & 26.3\% & 84.2\% & 0.0\% & 109 & 70 & 68 \\
\midrule
Q16 & How can code be made   understandable without extensive documentation? & DP & 26.3\% & 78.9\% & 5.3\% & 109 & 93 & 63 \\
Q160 & What is the status of   quantum-computing and what effect will it have on software development? & BEST & 26.3\% & 57.9\% & 10.5\% & 109 & 148 & 29 \\
Q106 & What factors affect a focus on   feature development instead of migrating towards new technologies or   architectures versus a focus on migration towards new technologies or   architectures as a part of feature development? & BEST & 25.0\% & 87.5\% & 0.0\% & 113 & 56 & 68 \\
Q24 & How can software development be   simplified in order to make it accessible for more people? & PROC & 25.0\% & 81.3\% & 12.5\% & 113 & 82 & 15 \\
Q157 & What impact does code quality   have on the ability to monetize a software service? & DP & 25.0\% & 80.0\% & 5.0\% & 113 & 84 & 67 \\
\midrule
Q13 & How can an inventory of code to   be re-compiled or re-linked be prepared when a configuration item is changed? & BEST & 25.0\% & 75.0\% & 0.0\% & 113 & 106 & 68 \\
Q54 & How does the use of a shell on   Unix - both in industry and within \ING - influence performance of software   developing teams? & BEST & 25.0\% & 62.5\% & 6.3\% & 113 & 142 & 55 \\
Q138 & What factors affect the   development- and operations (DevOps) tools used by system engineers versus   front-end engineers? & PROC & 25.0\% & 50.0\% & 0.0\% & 113 & 157 & 68 \\
Q51 & How does knowledge sharing of   backlog components affect team performance? & TC & 23.5\% & 70.6\% & 5.9\% & 119 & 121 & 59 \\
Q171 & Why is security by many   developers seen as 'not sexy'? & BUG & 23.1\% & 46.2\% & 7.7\% & 120 & 162 & 47 \\
\bottomrule
\end{tabular}
\end{table*}

\begin{table*}[h]
\label{tab:all_descriptive_questions6}
\small
\begin{tabular}{p{0.4cm}p{7.8cm}p{1.0cm}|P{0.8cm}P{1.2cm}p{0.7cm}|P{0.8cm}P{1.2cm}P{0.7cm}}
\toprule
 &  &  & \multicolumn{3}{c}{Percentages} & \multicolumn{3}{c}{Rank} \\
 & Question & Category & Essential & Worthwhile+ & Unwise & Essential & Worthwhile+ & Unwise \\ \midrule
Q158 & What is the data flow structure   of the application like upstream and downstream information of the   application? & DP & 22.2\% & 88.9\% & 0.0\% & 121 & 44 & 68 \\
Q40 & How do brightness and contrast   of monitors relate to human eyes and brain with regard to fatigue? & BEST & 22.2\% & 88.9\% & 0.0\% & 121 & 44 & 68 \\
Q99 & To what extent does test   automation affect delivery of software solutions? & TP & 22.2\% & 88.9\% & 0.0\% & 121 & 44 & 68 \\
Q78 & To what extent affects   commitment of an individual software engineer or a DevOps team to desired   time-lines the performance of a DevOps team and the quality of software? & PROD & 22.2\% & 77.8\% & 0.0\% & 121 & 95 & 68 \\
Q102 & What are good ways for software   engineers keep up to date with relevant technological developments? & BEST & 22.2\% & 66.7\% & 0.0\% & 121 & 128 & 68 \\
\midrule
Q27 & How can software solutions in   one common language be developed in a way that it is applicable to every   person, regardless of ones interest in software development? & CR & 22.2\% & 55.6\% & 33.3\% & 121 & 152 & 1 \\
Q101 & To what extent should   pen-testing be done within a team itself, or by a specialized pen-testing   team? & TP & 22.2\% & 44.4\% & 0.0\% & 121 & 165 & 68 \\
Q23 & How can software complexity best   be measured with regard to agility of the code base? & DP & 21.1\% & 84.2\% & 5.3\% & 128 & 70 & 63 \\
Q170 & Why do many developers focus on   the newest of the newest? Why don't they leave this to a small group in order   to use time and effort more efficient? & DP & 21.1\% & 47.4\% & 26.3\% & 128 & 161 & 3 \\
Q35 & How can we create an overview   from all NPA's (Microsoft Network Policy and Access Services) with their   authorizations on Windows server? & DP & 21.1\% & 42.1\% & 10.5\% & 128 & 167 & 29 \\
\midrule
Q105 & What delays are most common   inside development projects and what are the most common reasons for these   delays, and how such delays be avoided? & PROC & 20.0\% & 93.3\% & 6.7\% & 131 & 19 & 51 \\
Q124 & What factors affect platform   selection with regard to data being processed on it? & DP & 20.0\% & 90.0\% & 0.0\% & 131 & 30 & 68 \\
Q68 & To what extent affect different   branching mechanisms the performance of DevOps teams and the quality of code? & DP & 20.0\% & 90.0\% & 0.0\% & 131 & 30 & 68 \\
Q4 & Do distributed version control   systems offer any advantages over centralized version control systems? & BEST & 20.0\% & 90.0\% & 0.0\% & 131 & 30 & 68 \\
Q79 & To what extent affects data   science the quality of code or the performance of DevOps teams? & EQ & 20.0\% & 86.7\% & 0.0\% & 131 & 59 & 68 \\
\midrule
Q130 & What factors affect software   engineer productivity with regard to being a polyglot versus becoming an   expert in one language? & PROD & 20.0\% & 70.0\% & 10.0\% & 131 & 122 & 32 \\
Q62 & Is a static typed language   better than a dynamic typed language? & BEST & 20.0\% & 60.0\% & 0.0\% & 131 & 144 & 68 \\
Q61 & How well do time estimates   approximate the actual time taken to complete a project? & DP & 20.0\% & 55.0\% & 15.0\% & 131 & 154 & 12 \\
Q66 & Since there is a lot of   variation of methods used within \ING, what factors affect software delivery   with regard to the different software development methods that are used in   practice? & PROC & 18.8\% & 75.0\% & 6.3\% & 139 & 106 & 55 \\
Q21 & How can PL1 software code be   converted to Cobol code, while maintaining readability of the code in order   to simplify an application environment? & BEST & 18.2\% & 36.4\% & 18.2\% & 140 & 169 & 8 \\
\midrule
Q5 & Do distributed version control   systems offer any advantages over centralized version control systems? & PROD & 18.2\% & 18.2\% & 9.1\% & 140 & 171 & 36 \\
Q25 & How can software development   process simulations help to examine the impact of changes such as new   policies and changes in the way of working? & PROC & 17.6\% & 82.4\% & 0.0\% & 142 & 75 & 68 \\
Q149 & What factors affect the use of   machine learning in software development over a period of ten years? & DP & 16.7\% & 66.7\% & 16.7\% & 143 & 128 & 10 \\
Q46 & How do static code analysis   tools such as Fortify and Sonar influence the quality of software engineering   products? & BEST & 14.3\% & 85.7\% & 0.0\% & 144 & 64 & 68 \\
Q155 & What factors affect using   frameworks such as RIAF and BakerCatlogue as feature-rich monoliths versus   smaller programs with long-time support (LTS) versions? & DP & 14.3\% & 71.4\% & 0.0\% & 144 & 119 & 68 \\
\bottomrule
\end{tabular}
\end{table*}

\begin{table*}[h]
\label{tab:all_descriptive_questions7}
\small
\begin{tabular}{p{0.4cm}p{7.8cm}p{1.0cm}|P{0.8cm}P{1.2cm}p{0.7cm}|P{0.8cm}P{1.2cm}P{0.7cm}}
\toprule
 &  &  & \multicolumn{3}{c}{Percentages} & \multicolumn{3}{c}{Rank} \\
 & Question & Category & Essential & Worthwhile+ & Unwise & Essential & Worthwhile+ & Unwise \\ \midrule
Q128 & What factors affect running   individual software programs with regard to a dedicated versus a shared   environment? & EQ & 14.3\% & 71.4\% & 7.1\% & 144 & 119 & 49 \\
Q93 & To what extent do data   scientists affect the delivery of software solutions? & PROC & 14.3\% & 57.1\% & 0.0\% & 144 & 150 & 68 \\
Q156 & What factors affect working in   squads versus working in traditional project teams? & TC & 14.3\% & 42.9\% & 14.3\% & 144 & 166 & 13 \\
Q9 & Does a trade-off between existing   knowledge and usefulness of emerging technology affect the choice of a   programming language for an upcoming project? & PROC & 13.3\% & 66.7\% & 6.7\% & 149 & 128 & 51 \\
Q44 & How do risk management efforts   lead to less security related incidents and higher availability of systems? & BUG & 12.5\% & 75.0\% & 0.0\% & 150 & 106 & 68 \\
\midrule
Q3 & Debugging old code often is   complex; what factors affect the quality of legacy systems with regard to   debugging complexity, retention of developers, requirements, and   documentation? & PROC & 12,5\% & 68,8\% & 6,3\% & 150 & 124 & 55 \\
Q77 & To what extent affects changing   of requirements during development the delivery of software solutions? & PROC & 12.5\% & 68.8\% & 18.8\% & 150 & 124 & 7 \\
Q42 & How do dependencies in Java code   affect satisfaction of software engineers? & DP & 12.5\% & 62.5\% & 0.0\% & 150 & 142 & 68 \\
Q91 & To what extent affects the use   of a code generator the quality of software? & EQ & 12.5\% & 56.3\% & 0.0\% & 150 & 151 & 68 \\
Q117 & What factors affect installing   new infra-services - such as a server - with regard to configuration effort   needed, lead setup time, and cost? & PROC & 11.1\% & 88.9\% & 0.0\% & 155 & 44 & 68 \\
\midrule
Q81 & To what extent affects licensing   of tools the performance and productivity of DevOps teams? & DP & 10.5\% & 63.2\% & 0.0\% & 156 & 141 & 68 \\
Q86 & To what extent affects the   choice of a data modelling approach affects performance and productivity of a   DevOps team? & DP & 10.5\% & 57.9\% & 5.3\% & 156 & 148 & 63 \\
Q57 & How much time should be spent on  average on sharing knowledge within a software development team? & BEST & 10.0\% & 80.0\% & 10.0\% & 158 & 84 & 32 \\
Q167 & When does it make sense to   reinvent the wheel versus use an existing library? & BEST & 10.0\% & 60.0\% & 0.0\% & 158 & 144 & 68 \\
Q60 & How viable is it to use a formal  method such as model checking next to testing? & BEST & 10.0\% & 60.0\% & 0.0\% & 158 & 144 & 68 \\
\midrule
Q15 & How can be described in simple,   daily terms how software products run? & DP & 10.0\% & 50.0\% & 10.0\% & 158 & 157 & 32 \\
Q151 & What factors affect the use of   technical writers with regard to organization of writings and documentation   in order to increase trace-ability and readability? & DP & 9.1\% & 63.6\% & 9.1\% & 162 & 139 & 36 \\
Q110 & What factors affect data   analytics with regard to the use of external sources - such as market   research reports and follow market trends - and let individual teams handle   their local evolution? & PROC & 9.1\% & 63.6\% & 0.0\% & 162 & 139 & 68 \\
Q38 & How can we measure the time to   market of software solutions delivered within a department at \ING in order to   benchmark the performance of that department against others. & DP & 9.1\% & 54.5\% & 18.2\% & 162 & 155 & 8 \\
Q39 & How can Windows-server images be   created in order to facilitate testing within a continuous delivery pipeline? & DP & 9.1\% & 45.5\% & 27.3\% & 162 & 163 & 2 \\
\midrule
Q65 & Is there a correlation between   certification and developer effectiveness in the field? & PROD & 9.1\% & 45.5\% & 9.1\% & 162 & 163 & 36 \\
Q89 & To what extent affects the time   spent - in terms of full-time versus part-time - of a Scrum master the delivery   of software solutions? & PROC & 6.7\% & 20.0\% & 13.3\% & 167 & 170 & 14 \\
Q28 & How can the cost of data be   identified, in order to sign a price tag to data? & DP & 5.6\% & 50.0\% & 16.7\% & 168 & 157 & 10 \\
Q169 & Why are there so many open   source projects for the same function? & BEST & 5.6\% & 38.9\% & 11.1\% & 168 & 168 & 20 \\
Q49 & How does experience of software   developers and ops-engineers influence performance and quality of software   products? & BEST & 0.0\% & 66.7\% & 0.0\% & 170 & 128 & 68 \\
\midrule
Q72 & To what extent affects a clear separation between tests - such as unit test, integration test, end-to-end test - the delivery of software solutions? & TP & 0.0\% & 60.0\% & 10.0\% & 170 & 112 & 18 \\ 
\bottomrule
\end{tabular}
\end{table*}

\begin{table*}[t]
\caption{Statistically significant rating differences by demographics.}
\label{tab:demographics_discipline_all}
\small
\begin{tabular}{p{0.4cm}p{9.2cm}p{1.0cm}|p{1.5cm}|P{1.0cm}P{1.1cm}P{0.9cm}}
\toprule
 &  &  &  & \multicolumn{3}{c}{Discipline} \\
 & Question & Category & Response & Dev & Test & PM \\
 \midrule
Q2 & Are there practices of good software teams from the perspective of releasing software solutions into production? & PROC & Essential & \textbf{66.7\%} & 5.6 \% & 11.1\% \\
Q110 & What factors affect data analytics with regard to the use of external sources - such as market research reports and follow market trends - and let individual teams handle their local evolution? & PROC & Essential & \textbf{66.7\%} & 5.6 \% & 11.1\% \\
Q89 & To what extent affects the time spent - in terms of full-time versus part-time - of a Scrum master the delivery of software solutions? & PROC & Essential & \textbf{66.7\%} & 5.6 \% & 11.1\% \\
Q21 & How can PL1 software code be converted to Cobol code, while maintaining readability of the code in order to simplify an application environment? & BEST & Essential & \textbf{66.7\%} & 4.8 \% & 0.0\% \\
Q88 & \textit{To what extent affects the test ability of software code the quality of code?} & EQ & Essential & \textbf{68.4\%} & 0.0 \% & 0.0\% \\
Q95 & To what extent do dependencies on other teams affect team performance? & TC & Essential & \textbf{68.4\%} & 0.0 \% & 0.0\% \\
Q162 & What methods are most effective in preventing security related vulnerabilities or bugs from being introduced in software code? & BUG & Essential & \textbf{68.4\%} & 0.0 \% & 0.0\% \\
Q28 & How can the cost of data be identified, in order to sign a price tag to data? & DP & Essential & \textbf{72.7\%} & 0.0 \% & 0.0\% \\
Q97 & To what extent does documentation during software maintenance affect delivery of software solutions? & TP & Essential & \textbf{50.0\%} & 0.0 \% & 0.0\% \\
Q46 & How do static code analysis tools such as Fortify and Sonar influence the quality of software engineering products? & BEST & Essential & \textbf{36.6\%} & 0.0 \% & 27.3\% \\
\bottomrule
\\[-2pt]
\end{tabular}
\footnotesize\emph{}The demographic with the highest rating is highlighted in \textbf{bold}. Questions that are also in Table \ref{tab:essential_ranked} are shown in \textit{italics}.
\end{table*}

\begin{table*}[h]
\label{tab:demographics_role_all}
\small
\begin{tabular}{p{0.4cm}p{9.2cm}p{1.0cm}|p{1.5cm}|P{1.0cm}P{1.1cm}P{0.9cm}}
\toprule
 &  &  &  & \multicolumn{3}{c}{Management Role} \\
 & Question & Category & Response & Manager & Individual & Architect \\
 \midrule
Q2 & Are there practices of good software teams from the perspective of releasing software solutions into production? & PROC & Essential & 41.4\% & \textbf{44.8} \% & 6.9\% \\
Q153 & What factors affect trunk-based development - a source-control branching model, where developers collaborate on code in a single branch - with regard to quality of software code? & BEST & Essential & 22.6\% & \textbf{54.8} \% & 9.7\% \\
Q97 & To what extent does documentation during software maintenance affect delivery of software solutions? & TP & Essential & 10.0\% & \textbf{60.0} \% & 20.0\% \\
Q46 & How do static code analysis tools such as Fortify and Sonar influence the quality of software engineering products? & BEST & Essential & \textbf{69.2\%} & 15.4 \% & 0.0\% \\
\bottomrule
\\[-2pt]
\end{tabular}
\footnotesize\emph{}The demographic with the highest rating is highlighted in \textbf{bold}. Questions that are also in Table \ref{tab:essential_ranked} are shown in \textit{italics}. The role "Manager" includes the responses for "Manager" and "Lead".
\end{table*}

\twocolumn
\end{document}